\documentclass[useAMS,usenatbib]{mnras}
\usepackage{amsmath,fleqn,graphicx,amssymb,xcolor}
\usepackage{multirow}
\renewcommand{\[}{\begin{equation}}
\renewcommand{\]}{\end{equation}}
\def\i{{\rm i}}

\let\boldgrk=\gkvecten
\let\boldgrksc=\gkvecseven

\def\gkthing#1{{\mathchoice%
	{\hbox{{\boldgrk\char#1}}}
	{\hbox{{\boldgrk\char#1}}}
	{\hbox{{\boldgrksc\char#1}}}
	{\hbox{{\boldgrksc\char#1}}}}}

\def\vtheta{\gkthing{18}}

\def\vsigma{\gkthing{27}}

{\newif\ifnotend
\notendtrue
\def\veclist{ABCDEFGHIJKLMNOPQRSTUVWXYZabcdefghijklmnopqrstuvwxyz.}
\def\top#1#2.{#1}
\def\tail#1#2.{#2.}
\loop\expandafter\xdef\csname v\expandafter\top\veclist\endcsname%
{{\noexpand\bf\expandafter\top\veclist}}
\edef\veclist{\expandafter\tail\veclist}
\if\veclist.\notendfalse\fi\ifnotend\repeat}
{\newif\ifnotend
\notendtrue
\def\callist{ABCDEFGHIJKLMNOPQRSTUVWXYZ.}
\def\top#1#2.{#1}
\def\tail#1#2.{#2.}
\loop\expandafter\xdef\csname c\expandafter\top\callist\endcsname%
{{\noexpand\cal\expandafter\top\callist}}
\edef\callist{\expandafter\tail\callist}
\if\callist.\notendfalse\fi\ifnotend\repeat}

\def\ex#1{\left<#1\right>}

\def\d{{\rm d}}

\def\mnras{MNRAS}
\def\araa{AnnRA\&A}
\def\apj{ApJ}
\def\apjs{ApJS}
\def\apjl{ApJL}
\def\nat{Nat}
\def\aj{AJ}
\def\aap{A\&A}

\def\pasp{PASP}

\def\kpc{\,\mathrm{kpc}}

\def\kms{\,\mathrm{km\,s}^{-1}}

\def\masyr{\,\mathrm{mas\,yr}^{-1}}

\def\msun{\,{\rm M}_\odot}

\def\e{\mathrm{e}}

\def\fracj#1#2{{\textstyle{#1\over#2}}}

\def\eqrf#1{(\ref{#1})}

\title[Modelling the stellar halo with RR-Lyrae stars]
{Modelling the stellar halo with RR-Lyrae stars}

\author[Chengdong Li \& James Binney]{
  Chengdong Li$^1$\thanks{E-mail: chenglong.li@physics.ox.ac.uk} 
\& James Binney$^1$\thanks{E-mail:
  binney@physics.ox.ac.uk}\\   
  $^1$Rudolf Peierls Centre for Theoretical Physics, Clarendon Laboratory,
  Parks Road, Oxford, OX1 3PU, UK
}

\begin{document}
\maketitle

\begin{abstract}
A seven-parameter distribution function (DF) is fitted to $20\,000$ RR-Lyrae
stars for which only astrometric data are available. The observational data
are predicted by the DF in conjunction with the gravitational potential of a
self-consistent model Galaxy defined by DFs for the dark halo, the bulge and
a four-component disc. Tests of the technique developed to deal with missing
line-of-sight velocities show that adding such velocities tightens
constraints on the DF only slightly. The recovered model of the RR-Lyrae
population confirms that the population is flattened and has a strongly
radially biased velocity distribution. At large radii its density profile
tends to $\rho\sim r^{-4.5}$ but no power law provides a good fit inside the
solar sphere.  The model is shown to provide an excellent fit to the data
for stars brighter than $r=16.5$ but at certain longitudes it predicts too
few faint stars at Galactocentric radii $\sim20\kpc$, possibly signalling
that the halo is not axisymmetric. The DF is used to
predict the velocity distribution of BHB stars for which space velocities are
available. The $z$ components are predicted successfully but too much
anisotropy in the $v_Rv_\phi$ plane is expected.
\end{abstract}


\section{Introduction} \label{sec:intro}

Although its mass is relatively small, the stellar halo of our Galaxy is of
great interest for several reasons. Among these is the fact that it contains
many of the oldest stars in the Galaxy, and it is widely believed that
important insights into the Galaxy's history can be gleaned from the halo's
chemodynamical structure
\citep{Beea08,Belokurov2018,LancasterKoposov2019,MyeongVasiliev2019,BelokurovSanders2020}.
The stellar halo is also important as a tracer of the Galaxy's gravitational
field. Indeed, the kinematics of disc stars and cool gas have enabled us to
map the gravitational field in significant detail near the plane for several
kiloparsecs either side of the solar radius $R_0$
\citep{BinneyRAVE2014,Binney2015,ColeBinney,Bland-Hawthorn2016}, but the structure of the field
far from the plane and beyond $R\sim1.5R_0$ is not tightly constrained.
Available tracers of the field at these locations include tidal streams
\citep{KoRiHo09,Bovy2014:streams,Gibbons2016,Erkal2016,Koposov2019},
globular clusters \citep{Binney2017,VasilievGC2019} and dwarf spheroidal satellites
\citep{WiEv99}, and populations of halo stars
\citep{DeasonVasilyWyn2011MNRAS,DeasonVasilyJason2019}.

RR-Lyrae stars are probably the most useful population of halo stars because
they are quite luminous ($M_V\sim0.5$) so can be detected within a large
volume, and their distances can be obtained to
excellent precision from their period-luminosity relation. 
\cite{Sesar2017} used machine-learning to identify $45\,000$
RR-Lyrae stars in photometric data from the PanSTARRS1 survey \citep{Kaiser2010}.
Astrometric data for most of these stars are available in the second and
third data releases from the European Space Agency's satellite Gaia
\citep{GaiaCollaboration2018,GaiaEDR3}. Combining the proper-motion data from Gaia with distances
from the PanSTARRS photometry, we obtain precision five-dimensional data for
a population that traces the Galaxy's gravitational field.

Learning how to exploit such five-dimensional data  is an interesting and
potentially valuable exercise because through parallaxes and proper motions
Gaia provides five dimensional data for in excess a billion stars. To date
most studies of Galactic structure that use Gaia data restrict attention to
the $\la7$ million stars in the DR2 radial-velocity sample because for these
stars one has six-dimensional data. With such data, each star has a fairly
well defined location in phase space. Working with five-dimensional data is
much more challenging because we are ignorant of the line-of-sight velocities
$v_\parallel$. Here we develop a technique for managing our ignorance.

A population of tracer objects can be used to constrain the Galaxy's
gravitational field only to the extent that the latter can be approximated by
a constant field. Most halo objects are not expected to respond strongly to
the Galaxy's rotating bar or spirals, and the Galaxy's potential well deepens on a
timescale that is too long to invalidate the assumption of instantaneous
equilibrium. Within Galactocentric radii that are probed by  RR-Lyrae stars,
disequilibrium caused by the orbits of the Magellanic Clouds probably also
has too long a timescale to be
important. The same may not be true of the disequilibrium induced by the
Sagittarius dwarf galaxy, but a reasonable strategy seems to be to ignore the
Sagittarius dwarf in the first instance and then to use the resulting model
to investigate the impact that the Sgr Dwarf has on a population like that of
the RR-Lyrae stars.

For these reasons our scheme for exploiting the RR-Lyrae sample is to seek a
distribution function (DF) with these objects given the observations and a
time-independent and axisymmetric model of the Galaxy's gravitational field.
By Jeans theorem \citep{Jeans1915} the DF can be taken to be a function of
the isolating integrals of motion in the given gravitational field, and these
integrals are most conveniently taken to be three actions: $J_r$, which
quantifies te amplitude of radial excursions, $J_z$, which quantifies
oscillations either side of the Galactic plane, and $J_\phi$, which is the
component of angular momentum parallel to the Galaxy's assumed symmetry axis.
Hence we assume that the DF of the RR-Lyrae stars is an analytic function
$f(\vJ)$. The model gravitational field in which the RR-Lyrae stars are
assumed to move is that of a self-consistent model Galaxy that is defined by
a series of similar DFs, one for each of the principal mass-bearing
components of the Galaxy. Specifically, the dark halo, an (axisymmetric)
bulge and four disc components: a young disc, a middle-aged disc and an old
thin disc plus a high-$\alpha$ disc. The gravitational potential that these
DFs jointly generate is found iteratively in the manner originally described by
\cite{Binney2014} using the {\it AGAMA} package \citep{Vasiliev2019}, which
computes actions and gravitational potentials with remarkable efficiency, and
takes care of the complex book-keeping involved in self-consistent galaxy
modelling.

In this paper our focus is on determination of the DF of a sample of tracers
for which we have five-dimensional data. Hence we will consider many possible
DFs for the sample but use only one gravitational potential, which will arise
from, and be defined by, a single set of DFs for the Galaxy's mass-bearing
components. In a subsequent paper we will consider the extent to which data
for a tracer population such as the RR-Lyrae stars can be used to constrain
the DFs of the mass-bearing populations. 

The plan of this paper is as follows. Section~\ref{sec:DF} gives the
functional form of the DF we fit, Section~\ref{sec:fitting} summarises the
general theory of fitting DFs to data for individual stars and
Section~\ref{sec:compute} outlines the numerical procedure for evaluating
distribution functions.  Section~\ref{sec:gaia_error} explains how
correlations between parallax and proper motion in Gaia data can be used to
refine proper-motion data when an accurate spectrophotometric distance is
available.  Section~\ref{sec:obs} defines the sample we use and its selection
function.  Section~\ref{sec:test} demonstrates that MCMC exploration of model
space recovers from realistic data the parameters of a DF to within the
predicted uncertainties.  Section~\ref{sec:realdata} reports the results of
applying the technique to the RR-Lyrae data. Section~\ref{sec:quality}
investigates the success with which the recovered DF fits the observational
data and shows that the fit is good when a plausible modification is made to
the selection function.  Section~\ref{sec:bhbtest} explores the success of
the model in predicting data for another tracer of the stellar halo, namely
blue-horizontal-branch (BHB) stars. Section~\ref{sec:prior} compares our
technique to that of \cite{Wegg2019}, who modelled very similar data in a
different way, and compares our model with those previously derived.  Finally
Section~\ref{sec:conclusion} sums up and indicates directions for future
work.

\section{Modelling technique}\label{sec:theo}

We model the RR-Lyrae population as a massless component of a self-consistent
model Galaxy previously constructed using the {\it AGAMA} software library.
The assumption of negligible mass greatly simplifies the process of fitting a
model to data and is an excellent approximation given that the mass of the
entire stellar halo, of which the RR-Lyraes are a part, is only $\sim$
$4-7\times~10^8~M_{\odot}$ \citep{Bland-Hawthorn2016}.

In {\it AGAMA} a model of this type is specified by an `ini-file', and the relevant file is
available in the online material. The distribution functions of the
mass-bearing components are by far the most important part of the
specification. For our model these are the dark halo, the bulge and a
four-component stellar disc -- the disc's components are young, middle-aged
and old discs of low-$\alpha$ stars, and a disc of old, high-$\alpha$ stars.
 The details of the disc model will be presented in Binney \& Vasiliev
(in preparation) and Li \& Binney (in preparation).
The distribution function of each component is a specified function $f(\vJ)$
of the action integrals. The model's potential
derives from the above distribution functions plus a gas disc. This has a
mass of $1.04\times10^{10}\msun$ with a surface density that declines exponentially with
scale length $5\kpc$ beyond the Sun but interior to the Sun has a depression
in surface density of the type introduced by \cite{DehnenBinney98}.

The DFs of the dark halo, the bulge and the stellar halo are of the `softened
double-power-law' type described below. The DFs of the disc components are of
the `exponential' type that is defined by Binney \& Vasiliev (in
preparation). Complete data for reproducing the model with {\it AGAMA} are
contained in the file {\tt m.ini} that can be found in the online addendum.

\subsection{DF for spheroidal populations}\label{sec:DF}

The DF of the RR-Lyrae population is based on the double-power law distribution
function introduced by \citet{Posti2015}
\begin{equation}\label{eq:DF}
    f(\vJ)=\frac{M}{(2\pi J_{0})^{3}}\frac{(1+[J_{0}/h_{\vJ}]^{\gamma})^{\alpha/\gamma}}
    {(1+[g_{\vJ}/J_{0}]^{\gamma})^{\beta/\gamma}}\e^{-(g_{J}/J_{\rm
    cut})^{\delta}}.
\end{equation}
Here the normalisation constant $M$ sets the mass of the population while 
$h(\mathbf{J})$ and $g(\mathbf{J})$ are  two linear combinations of
the actions
\begin{equation}
    \begin{aligned}
    &h(\mathbf{J}) = (3-h_{\phi}-h_{z})\,J_{r}\,+\,0.7(h_{z}\,J_{z}+h_{\phi}\,
    \left|J_{\phi}\right|),\\
    &g(\mathbf{J}) = (3-g_{\phi}-g_{z})\,J_{r}\,+\,0.7(g_{z}\,J_{z}+g_{\phi}\,
    \left|J_{\phi}\right|).
    \end{aligned}
\end{equation}
The function $h(\vJ)$ defines the behaviour of the DF as $\vJ\to0$ and
therefore structure the population at small radii, while the function
$g(\vJ)$ defines the behaviour of the DF as $\vJ\to\infty$ and therefore
structures the population at large radii. The dimensionless parameter
$\alpha$ sets the slope of the power law that $h(\vJ)$ generates, while the
slope of the outer power law is set by $\beta$. The parameter $\gamma$
controls the sharpness of the transition from one regime to the other.  Since
early experiments showed that our data barely constrain $\gamma$, we
set it to unity. The dimensionless coefficients $h_\phi$ and $h_z$ determine
the degree of velocity anisotropy at small radii: increasing $h_\phi$ above
unity introduces radial bias in the sense of causing the radial velocity
dispersion $\sigma_R$ to exceed the azimuthal velocity dispersion
$\sigma_\phi$. Similarly, increasing $h_z$ above unity decreases the vertical
velocity dispersion $\sigma_z$ and thus tends to flatten the population.

The parameter $J_0$ in the DF \eqrf{eq:DF} sets the break radius at which the
flatter density profile generated by $h(\vJ)$ goes over to the steeper
profile generated by $g(\vJ)$. The parameter $J_{\rm cut}$ in the DF
\eqrf{eq:DF} sets the radius at which the outer power-law (which may be
incompatible with finite mass) gives way to a steep downturn in the density.
The sharpness of the downturn is set by the parameter $\delta$.  $J_{\rm
cut}$ is set to $20\,000\kpc\kms$ for the dark halo, $280\kpc\kms$ for the
bulge and $100\,000\kpc\kms$ for the stellar halo. This last value is too
lage to have a significant influence on the stellar halo at radii probed by
observational data.
The parameters that can be materially constrained by
observational data are $\alpha$, $\beta$, $h_\phi$, $h_z$, $g_\phi$, $g_z$ and
$J_0$.

The double-power-law DF \eqrf{eq:DF} generates a cusped density distribution.
Since a cusp is both physically and observationally  problematic, following
\cite{ColeBinney} we
eliminate it by dividing the DF \eqrf{eq:DF} by
\begin{equation} \label{eq:soft}
S(x)=    [x^2-\beta x+1]^{\alpha/2}
\end{equation}
with $x=J_c/h(\vJ)$. Then when $|\vJ|\ll J_c$ and $x\gg1$ the divergence of
$S$ cancels the divergence of the DF caused by $h(\vJ)$ tending to zero,
while when $|\vJ|\gg J_c$, $S$ tends to unity and $f(\vJ)$ is unaffected by
the division. The parameter $\beta$ is set automatically to ensure that
dividing by $S$ does not change the total mass of the component. $J_c$ is set
to $300\kpc\kms$ for the dark halo and $50\kpc\kms$ for the bulge. The value
adopted for the stellar halo is immaterial because this halo contributes
negligibly to the gravitational potential and the observational data do not
penetrate close enough to the centre to be  affected by plausible values of
$J_c$. For simplicity, we adopted $J_c=0$.

\subsection{Fitting the DF to data}\label{sec:fitting}

The parameters of the DF are constrained by exploring the probability of the
survey data as a function of the DF's parameters following the scheme described by
\citet{McMillan2012,McMillan2013}.

 The required probability of the data given a model is a product of
probabilities for individual stars. The probability that a randomly chosen
survey star is located within the phase-space volume $\d^6\vw$ around
the phase-space location $\vw=(\vx,\vv)$
is
\[
P(\vw|f\hbox{ \& in survey})\,\d^6\vw={S(\vw)f(\vw)\over P_{\rm S}}\,\d^6\vw,
\]
where $S(\vw)$ is the probability that a star located at $\vw$ will be
captured by the survey (because it is bright enough and appears in an
observed region of the sky) and $f(\vw)$ is the distribution function
normalised such
that $\int\d^6\vw\,f=1$. The denominator
\[
P_{\rm S}=\int\d^6\vw\,S(\vw)f(\vw)
\]
is the probability that a star
randomly chosen from the population will appear in the catalogue. It
ensures that $P(\vw|f\hbox{ \& in survey})$ is a correctly normalised probability
density. 

On account of observational errors, we should maximise not $P(\vw|f\hbox{ \& in
survey})$ but the related probability density
\[
P_{\rm o}(\vw)=\int\d^6\vw'\,G(\vw-\vw',\vK)P(\vw'|f\hbox{ \& in survey})
\]
that the catalogue will list a star at $\vw$ that has true phase-space
location $\vw'$. Here we assume that the distribution of observational
errors $G$ is a multi-variate Gaussian with kernel $\vK$:
\[
G(\vw,\vK)=\sqrt{|\vK|\over(2\pi)^n}\exp(-\fracj12\vw^T\cdot\vK\cdot\vw),
\]
where $n=\hbox{dim}(\vw)$.
Thus $f$ should be chosen to maximise
\[
P_{\rm o}(\vw)={1\over P_{\rm S}}\int\d^6\vw'\,G(\vw-\vw',\vK)
S(\vw')f(\vw').
\]

In practice it is convenient to work with sky coordinates, which do not
comprise a system of canonical coordinates for phase space. Specifically, we
use Galactic longitude and latitude $\ell,b$, distance $s$, the proper
motions $\mu_\alpha=\dot\alpha\cos\delta$ and $\mu_\delta$ and the
line-of-sight velocity $v_\parallel$.  Forming these into the vector
$\vu=(\alpha,\delta,s,\mu_\alpha,\mu_\delta,v_\parallel)$, we have that the element of
phase-space volume $\d^6\vw$ is related to $\d^6\vu$ by
\citep[e.g][]{McMillan2012}
\[
\d^6\vw=s^4\cos\delta\,\d^6\vu,
\]
so
\[
P_{\rm o}(\vu)={1\over P_{\rm S}}\int\d^6\vu'\,s^{\prime 4}\cos b'\,G(\vu-\vu',\vK)
S(\vw')f(\vw'),
\]
where $\vw'$ is understood to be a function of $\vu'$. One advantage of
working with sky coordinates is that the matrix $\vK$ then simplifies. Most
of its off-diagonal elements vanish and $K_{\alpha\alpha}$ and
$K_{\delta\delta}$ become
very large because the sky positions of stars have negligible uncertainty.
This being so $G$ may be approximated by the product of a $4\times4$ matrix
$\widetilde\vK$ and Dirac $\delta$-functions in $\alpha-\alpha'$ and
$\delta-\delta'$ so
the integrals over these coordinates can be trivially executed. In Gaia data
there are significant correlations between the errors in distance $s$ and the
proper motions, and in Section~\ref{sec:gaia_error} below we take advantage
of this correlation. Otherwise we neglect correlations by
approximating $\widetilde\vK$ by a diagonal matrix,\footnote{ When $s$
is photometric,
$\widetilde\vK$ has just a pair of non-zero off-diagonal elements,
$K_{\alpha\delta}=K_{\delta\alpha}$. These are small because the long axis of the probability
distribution that $\vK$ describes tends to be aligned with the right-ascension
axis. In future work we will take these off-diagonal terms fully into
account, but we do not expect them to have a significant impact on the
results.}
\[
\widetilde\vK=\hbox{diag}\big[\epsilon_s^{-2},\epsilon_{\mu_\alpha}^{-2},
\epsilon_{\mu_\delta}^{-2},\epsilon_{v_\parallel}^{-2}\big].
\]

Conversion of heliocentric coordinates into phase-space coordinates requires
knowledge of the Sun's Galactocentric position and velocity.  We use
Galactocentric Cartesian coordinates $(x,y,z)$ in which the Sun's position
vector is $ (-8.2, 0, 0)$. Heliocentric distances are denoted by $s$ and
Galactocentric distances by $r$.  From \cite{Schoenrich2012} we take the Sun's
Galactocentric velocity $\vV_\odot$ to be
\[
\vV_\odot=(U,V,W)=(11.1, 250.24, 7.25)\kms.
\]

\subsection{Evaluating the probabilities}\label{sec:compute}

Evaluation of the quality of a model requires execution of two distinct
numerical tasks. One is computation of $P_{\rm S}$  by integrating the
product of the DF and the survey's selection function through phase space.
Introducing angle-action variables $(\vtheta,\vJ)$,
we have
\[
P_{\rm S}=\int\d^3\vJ\,f(\vJ)\int\d^3\vtheta\,S(\vJ,\vtheta).
\]
Following \cite{McMillan2013} we execute this integral by the Monte-Carlo
principle:
\[
\int \d x\,f(x)\simeq{1\over N}\sum_{i=1}^N{f(x_i)\over f_{\rm s}(x_i)},
\]
where the points $x_i$ are randomly sampled according to the probability
density $f_{\rm s}$. We take $f_{\rm s}$ to be a function of $\vJ$ only, so
we can write
\[
P_{\rm S}\simeq{1\over N}\sum_i^N{f(\vJ_i)\over f_{\rm
s}(\vJ_i)}S(\vtheta_i,\vJ_i).
\]
Poisson noise is minimised if $f(\vJ)/f_{\rm s}(\vJ)\simeq1$, and in our case
this can be achieved by taking $f_{\rm s}$ to be the first DF we try. If the
coordinates $(\vtheta_i,\vJ_i)$ of the points that sample $f_{\rm s}$ and the
resulting values of the ratio $S(\vtheta_i,\vJ_i)/f_{\rm s}(\vJ_i)$ are
stored, the quality of any subsequently proposed DF can be computed cheaply
merely by evaluating it at the $\vJ_i$.

The other numerical task is evaluation of a four-dimensional integral through
the error ellipsoid around each data point. We evaluate this by a
Gauss-Hermite algorithm. When $v_\parallel$ has been measured, we obtain from
the {\it Python} utility {\tt
quadpy}\footnote{https://github.com/nschloe/quadpy} $n=193$ 4-dimensional
locations $\vx_i$ and weights $A_i$ for the weight function $\e^{-|\vx|^2}$ such
that for any function $g(\vx)$
\[
\int\d^4\vx\,\e^{-|\vx|^2}g(\vx)\simeq\sum_i^nA_ig(\vx_i).
\]
Then
setting  $h(\vu)=s^4S(\vu)f(\vu)$ and
\[
\vu_i=\vu+\surd2\sum_{\alpha=\ell,b,s,\parallel}\epsilon_\alpha
x_{i\alpha}\ve_\alpha,
\]
where $\ve_\ell=(1,0,0,0)$, $\ve_b=(0,1,0,0)$, etc,
we have that
\[
P_{\rm o}(\vu)\simeq{\cos b\over\pi^2 P_{\rm S}}\sum_i^nA_i h(\vu_i).
\]

A more complex procedure is required when $v_\parallel$ has not been measured,
for then the error ellipsoid becomes a section of a four-dimensional
cylinder, the cross-sections of which are three-dimensional ellipsoids
spanned by the measured quantities $(s,\mu_\alpha,\mu_\delta)$. We need to integrate
through only that part of the cylinder for which the Galactocentric speed $v$
is less than the escape speed because the DF vanishes for
$v(v_\parallel)>v_{\rm esc}$. The strategy we adopted is to obtain from {\tt
quadpy} $n=77$ locations $\vx_i$ and weights $A_i$ for three-dimensional integration with weight
function $\e^{-|\vx|^2}$. Then with $G(\vu|\vK)$ now denoting a
three-dimensional Gaussian distribution, we have at each $i$ that
\begin{align}
P_{\rm o}
&={\cos b\over V_0P_{\rm S}}
\int\d^3\vu\,G(\vu-\vu'|\vK)\int_{v_-}^{v_+}\d v_\parallel\,h(\vu,v_\parallel)\cr
&={\cos b\over V_0\pi^{3/2}P_{\rm S}}
\sum_{n=1}^{77} A_i \int_{v_-}^{v_+}\d v_\parallel\,h_i(v_\parallel),
\end{align}
where $V_0$ is a normalising velocity, $v_\pm$ are the values of $v_\parallel$  at which $v=v_{\rm esc}$ and 
\[
h_i(v_\parallel)=h\Big(\vu_i+\surd2\!\!
\sum_{\lambda=\alpha,\delta,s}\!\!\epsilon_\lambda x_{i\lambda}\ve_\lambda
+v_\parallel\ve_\parallel\Big).
\]
The integral over $v_\parallel$ was executed by Gauss-Legendre
integration with unit weight function using $20$ integrand
evaluations.\footnote{The
overall count of $77\times20=1540$ evaluations of the DF for each star
compares unfavourably with the 193 evaluations required with measured
$v_\parallel$. To devise a more economical scheme one should avoid splitting
the integral into three- and one-dimensional parts.} 
The constant $V_0$ is
the same for all stars and DFs, so it plays no role in the optimisation and
can be set to unity.

\subsection{Refining the proper motions}\label{sec:gaia_error}

Since the {\it Gaia} data-reduction
pipeline decomposes the wiggly path of each star across the sky into linear
proper motion and elliptical parallactic motion, its output is characterised
by significant correlations between proper motion and parallax. From the
photometric data we have rather precise distances for the RR-Lyrae stars and we can
use these distances and the correlation coefficients in EDR3 to obtain more
accurate and less uncertain proper motions. 

Let $\delta\mu=\mu-\overline{\mu}$ and
$\delta\varpi=\varpi-\overline{\varpi}$ be offsets in proper motion (in
either sky coordinate) and
in parallax from the central EDR3 values, and define
\[\label{eq:Douglas2}
\delta\vu\equiv\left(
\begin{matrix}\delta\mu\cr\varpi_0-\overline{\varpi}\end{matrix}\right),
\]
where $\varpi_0$ is the photometrically determined parallax.
Then the probability distribution of
$\delta\mu$ is
\begin{align}\label{eq:Douglas1}
\d&
P(\delta\mu|\varpi_0)={\exp\big(-\fracj12\delta\vu^T\cdot\vK\cdot\delta\vu)\big]\,\d\delta\mu
\over\int_{-\infty}^\infty\d\delta\mu\,\exp\big(-\fracj12\delta\vu^T\cdot\vK\cdot\delta\vu)\big]}\cr
&={\exp\big[-\fracj12K_{\mu\mu}\big(\delta\mu+\delta\varpi_0K_{\varpi\mu}/K_{\mu\mu})^2\big]\d\delta\mu\over
\int_{-\infty}^\infty\d\delta\mu\,\exp\big[-\fracj12K_{\mu\mu}\big(\delta\mu+\delta\varpi_0K_{\varpi\mu}/K_{\mu\mu})^2\big]
},\phantom{X}
\end{align}
 so $\delta\mu+\delta\varpi_0K_{\varpi\mu}/K_{\mu\mu}$ is a Gaussianly
distributed  random variable with zero mean and variance $1/K_{\mu\mu}$.
Hence
\begin{align}
\ex{\delta\mu}+\delta\varpi_0{K_{\varpi\mu}\over K_{\mu\mu}}&=0\cr
\ex{\big(\delta\mu-\ex{\delta\mu}\big)^2}&={1\over K_{\mu\mu}}.
\end{align}
Now we have a new expectation value for $\mu$
\[
\ex{\mu}=\overline{\mu}+\ex{\delta\mu}=\overline{\mu}-\delta\varpi_0{K_{\varpi\mu}\over
K_{\mu\mu}}
\]
and a new variance for $\mu$
\[
\ex{\big(\mu-\ex{\mu}\big)^2}=\ex{\big(\delta\mu-\ex{\delta\mu}\big)^2}
={1\over K_{\mu\mu}}.
\]
The matrix $\vK$ is the inverse of the correlation matrix
$\overline{u_iu_j}$, so 
\[
K_{\mu\mu}={\overline{\delta\varpi^2}\over|\vK|}\hbox{ and }
K_{\varpi\mu}=-{\overline{\delta\varpi\delta\mu}\over|\vK|},
\]
where
\[
|\vK|=\overline{\delta\mu^2}\overline{\delta\varpi^2}-\overline{\delta\varpi\delta\mu}^2.
\]
\begin{figure}
	\includegraphics[width=\columnwidth]{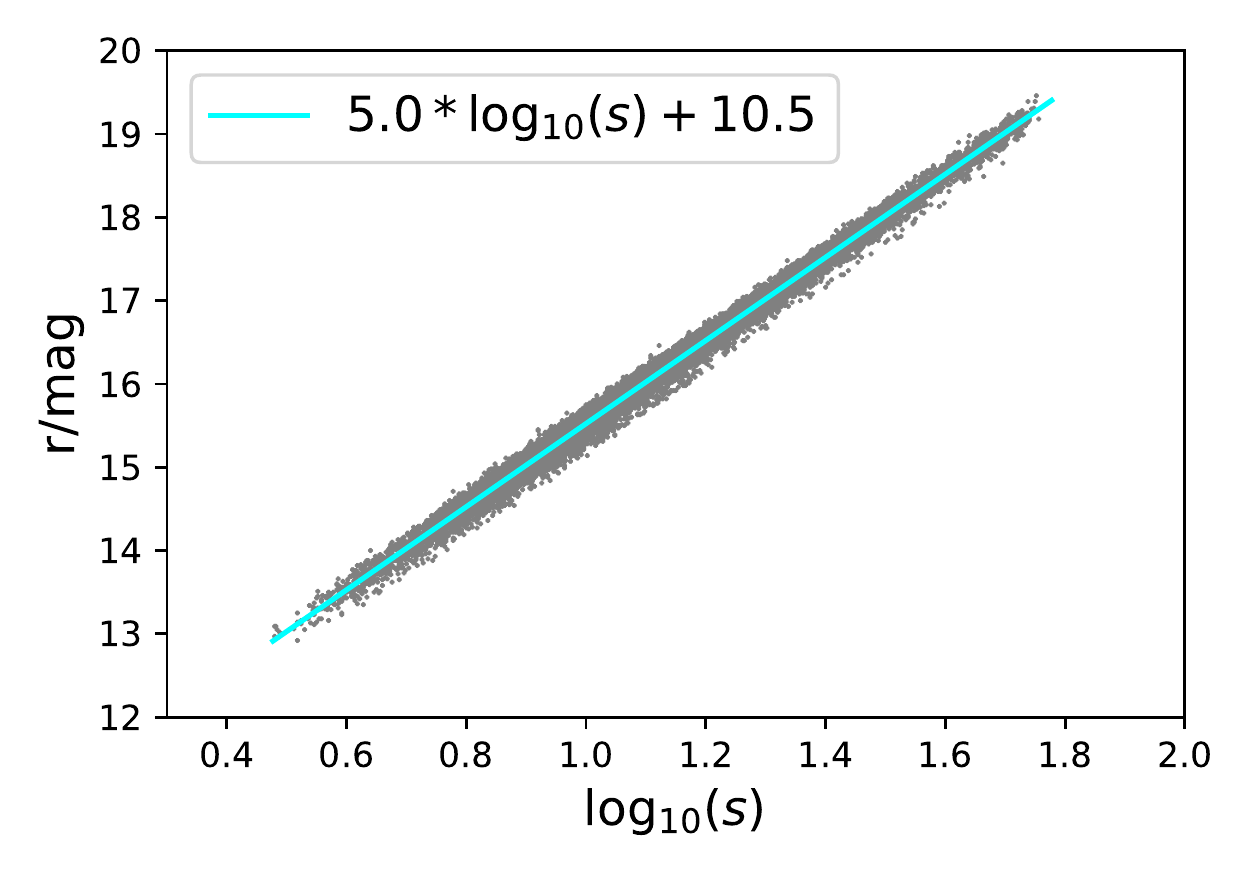}
    \caption{The correlation between $r$-band magnitude and distance for 
      RR-Lyrae stars. The cyan line shows the linear fit employed.}
    \label{fig:rf}
\end{figure}
\begin{figure*}
	\includegraphics[width=.8\hsize]{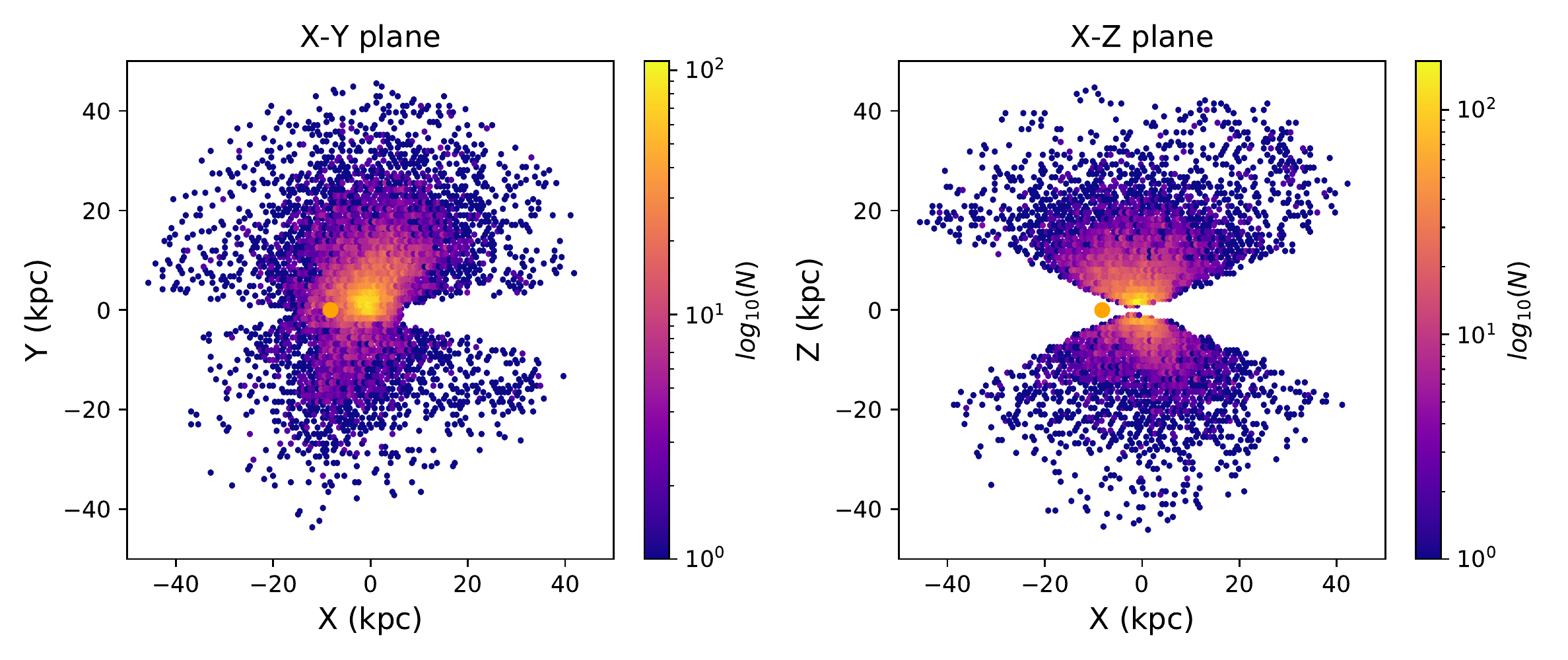}
    \caption{The spatial distribution of the RR-Lyrae sample showing the
    impact of  the selection function.}
    \label{fig:pos}
\end{figure*}

\begin{figure}
	\includegraphics[width=\columnwidth]{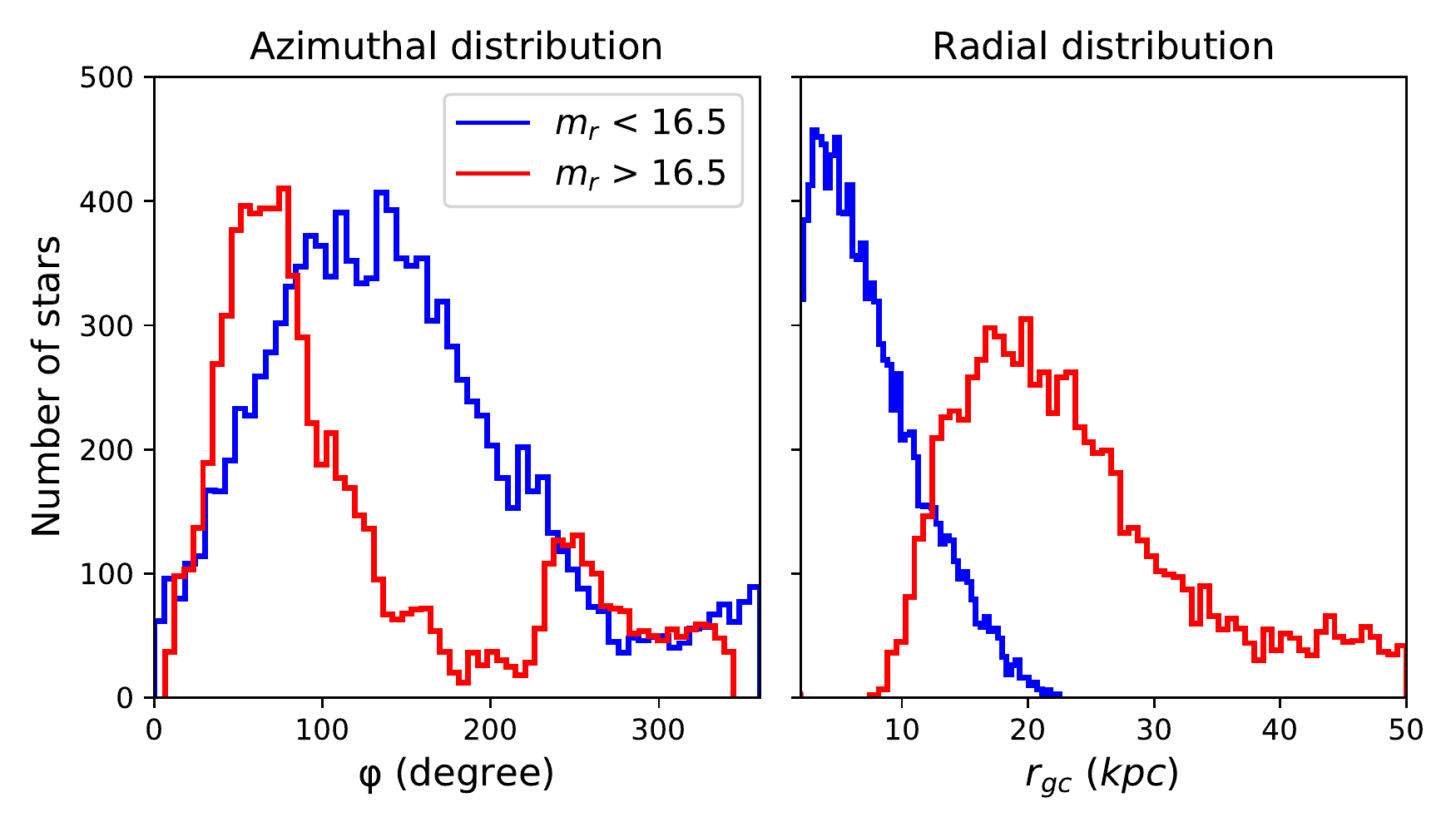}
    \caption{The azimuthal (left) and radial distributions of the stars. The
blue and red curves are for stars brighter or fainter than $r=16.5\,$mag,
respectively.}
    \label{fig:rphi}
\end{figure}

\section{Observational data}\label{sec:obs}

\subsection{The RR-Lyrae sample}

Our sample derives from the $3\pi$ section of the PanSTARRS1 (PS1) survey
\citep{Kaiser2010}. \citet{Sesar2017} used a machine-learning method to
identify RR-Lyrae stars in the PanSTARRS1 survey, and determined the purity
and completeness of the sample. We set their `threshold on score' to $0.5$
to obtain a sample of RR-Lyrae stars of $0.95$ purity and $0.97$
completeness for the stars at $\sim 18.5\,$mag in the $r$-band, which
corresponds to $40\kpc$ from the Sun \citep{Sesar2017}.  \citet{Sesar2017}
conclude that the distances, from the period-luminosity relation, have
uncertainties $\sim3\,$per cent.

We used astrometry from the \textit{Gaia} mission's early third data release
\citep{GaiaEDR3}. The typical uncertainties for parallax and
proper motion are $0.07\,$mas and $0.07\masyr$ at $G=17\,$mag and
$0.5\,$mas and $0.5\masyr$ at $G=20\,$mag for 5-parameter solutions
\citep{GaiaEDR3}.

\begin{figure*}
	\includegraphics[scale=0.45]{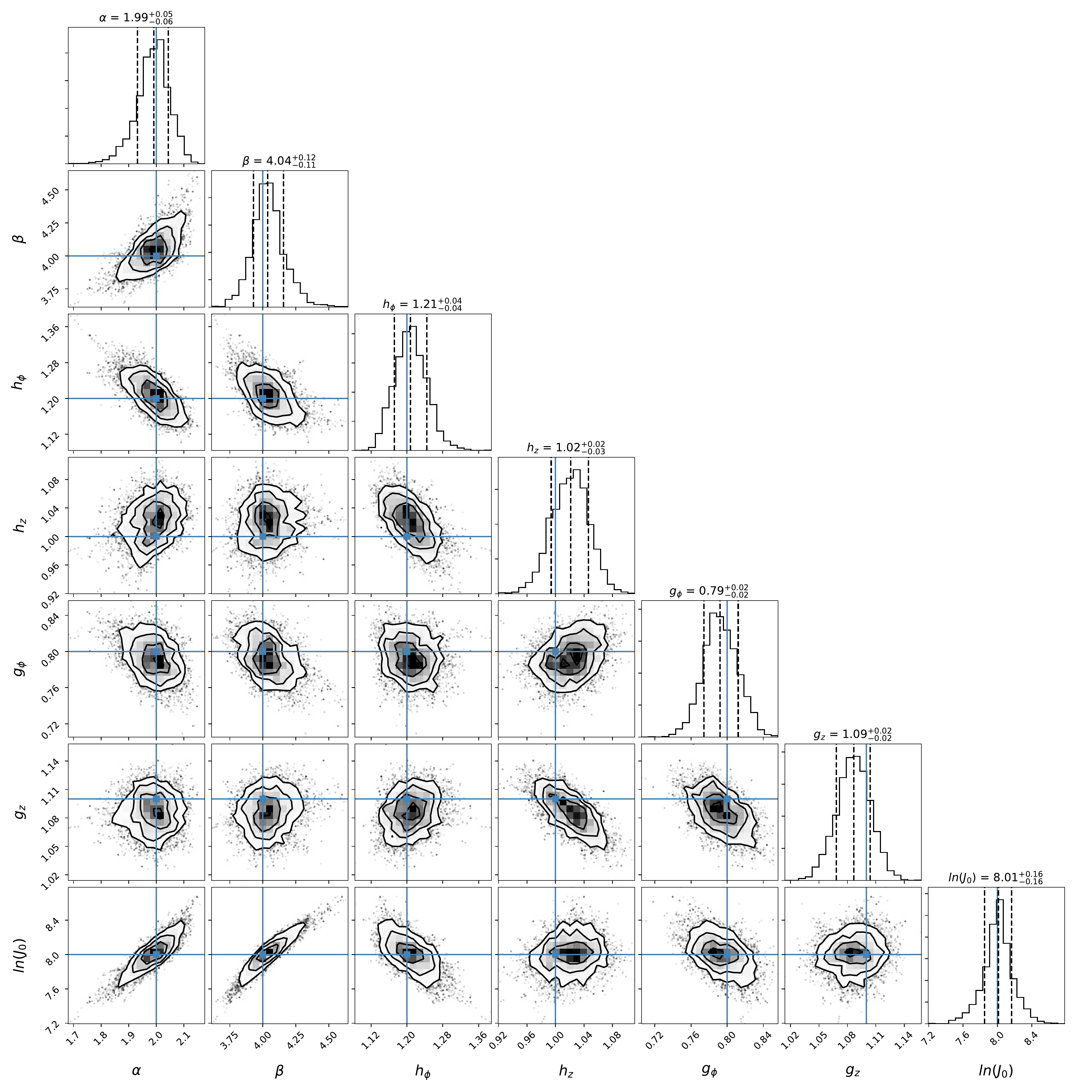}
    \caption{The posterior probability distributions and correlations of the
    parameters
    when the DF is fitted to  mock data that includes line-of-sight
    velocities. The dashed lines mark the 1-$\sigma$ uncertainty while the
    blue lines mark the true value of the parameter.}
    \label{fig:mock_6d}
\end{figure*}

\subsection{Selection function}\label{sec:selection}

\citet{Sesar2017} derived the selection function of the RRab stars in PanSTARRS1.
It is almost a constant at $\sim90\,$per cent for $r$-band magnitude
$r_F<20$, being well fitted by the Fermi-Dirac function  
\begin{equation}\label{eq:selectionfunction}
    S(r_F)={L\over \exp[k(r_F+x_0)]+1}.
\end{equation}
where $L=0.91$ represents the peak value of the selection function, $k=4.1$
is the sharpness and $x_0=20.6$ denotes the magnitude at which the
completeness drops to $50\,$per cent.  Given the almost fixed absolute
magnitude of these stars, $r_F$ is a function of distance, namely
\begin{equation}\label{eq:rF}
r_F=10.5+5\log_{10}(s/\!\kpc),
\end{equation}
which is based on the linear fit of the r-band apparent magnitude
with the distance shown in Figure~\ref{fig:rf}. Substituting equation
\eqrf{eq:rF} into equation \eqrf{eq:selectionfunction} we obtain the
selection function as a function of distance.

In view of the latitude of Hawaii, we consider only stars with $\delta>
-29^{\circ}$.  To avoid the contamination by the bulge and to avoid stars
with large proper-motion errors we restrict the sample to stars with
Galactocentric distances $r$ in the range $1.5\kpc<r<50\kpc$.  To avoid the
region of high extinction near the Galactic plane and to ensure the sample is
dominated by halo stars, we adopt the double cut
\begin{equation}\label{eq:cutangle}
	\frac{|z|}{\sqrt{x^{2}+y^{2}}}>\tan{20^{\circ}},\quad 
    |b|>10^{\circ}.
\end{equation}

Given that we are fitting to the data a smooth, axisymmetric model of the
stellar halo, we try to exclude stars located in over-dense regions. First,
we remove all  stars that lie within 10 half-light radii of a globular cluster
according to the method in \citet{Wegg2019}. We use the \citet{Harris1996}
catalogue of globular clusters and adopt $2^\prime$ when no half-light
radius is listed. 

Then we remove stars located in identified halo streams.  First, we remove
stars at $s>15\kpc$ that lie near the plane of the Sgr stream as defined by
\citet{Majewski2003}. Since stars in the trailing arm of Sgr are spread more
widely either side of the Sgr plane than stars in the leading arm, at $x<0$
we delete stars that lie within $15^\circ$ of the plane, while at $x>0$ we
cut stars that lie within $10^{\circ}$ of the plane. The transformation of
equatorial coordinates to Sgr coordinates is taken from
\citet{Belokurov2014}.  We also cut stars that belong to other streams
according to Tables 3 and 4 in \citet{Yang2019}.

Finally we cross match the RR-Lyrae catalogue with \textit{Gaia} EDR3 to
within a radius of $0.5^{\prime\prime}$ and remove stars without a
measurement of proper motion or $\tt{astrometric\_excess\_noise\_sig} > 10$.
The sample then comprises $19\,154$ RRab stars. The fraction of stars in each
of 25 bins in Galactocentric radius  that is successfully cross-matched to
EDR3 is noted and the selection function that will be applied to models
is updated by being multiplied by these fractions.

Figure~\ref{fig:pos} shows the distribution of the final sample, projected
onto the Galactic plane in the left panel and onto the meridional $(R,z)$
plane in the right panel. The Sun's location is marked by an orange dot.
Figure~\ref{fig:rphi}, shows the azimuthal and
radial distribution of the bright ($r<16.5$ blue curves) and faint stars in the
sample. Naturally very few faint stars appear within $10\kpc$ of the Galactic
centre.

\begin{table*}
	\centering
	\caption{Summary of all tests. The second and third rows give the
	means and 1-$\sigma$ uncertainties of the posterior distributions
	from  a single catalogue. The fourth and fifth rows show the means
	and standard deviations of the means that were extracted from ten different
	realisations of the same DF. The units of $J_0$ anr $\!\kpc\kms$.}
	\label{tab:finalcom}
	\begin{tabular}{llccccccc} 
		\hline
&		&$\alpha$ & $\beta$ & $h_{\phi}$ & $h_{z}$ & $g_{\phi}$
		& $g_{z}$ & $\ln{J_{0}}$\\
&		Input Value&2.0 & 4.0 & 1.2 & 1.0 & 0.8 & 1.1 & 8.0\\
\hline
1 catalogue&	6D results&$1.99^{+0.05}_{-0.06}$ & $4.04^{+0.12}_{-0.11}$ & $1.21^{+0.04}_{-0.04}$ & $1.02^{+0.02}_{-0.03}$ & $0.79^{+0.02}_{-0.02}$ & $1.09^{+0.02}_{-0.02}$ & $8.01^{+0.16}_{-0.16}$\\
	&	5D results& $2.00^{+0.06}_{-0.07}$ & $4.14^{+0.24}_{-0.17}$ & $1.22^{+0.05}_{-0.05}$ & $1.02^{+0.03}_{-0.03}$ & $0.78^{+0.03}_{-0.03}$ & $1.08^{+0.02}_{-0.02}$ & $8.11^{+0.25}_{-0.25}$\\
\hline
10 catalogues&	6D results& $2.04\pm0.05$ & $4.24\pm0.21$ & $1.19\pm0.03$ & $1.00\pm0.02$ & $0.78\pm0.02$ & $1.11\pm0.02$ & $8.27\pm0.21$\\
	&	5D results& $2.06\pm0.07$ & $4.40\pm0.32$ & $1.19\pm0.04$ & $1.01\pm0.02$ & $0.76\pm0.03$ & $1.09\pm0.02$ & $8.37\pm0.32$\\
		\hline
	\end{tabular}
\end{table*}

\section{Tests on mock data}\label{sec:test}

We tested our ability to recover the DF of the RR-Lyrae stars from realistic
data as follows. We generated ten sets of mock data by instructing {\it
AGAMA} to randomly sample a population of RR-Lyrae stars with a known DF --
the top row of Table~\ref{tab:finalcom} gives the
parameters of this DF. The {\it astropy} package \citep{Whelan2018} was used to
convert the Cartesian coordinates $(\vx,\vv)$ produced by {\it AGAMA} to
observational variables
$\vu=(\alpha,\delta,s,\mu_{\alpha},\mu_{\delta},v_\parallel)$ . Each mock
star was then included in, or excluded from, the mock catalogue by applying
the selection function described in Section \ref{sec:selection}. Finally, the
data were scattered by observational errors appropriate to the star's
apparent magnitude. The ten resulting mock catalogues, each with
 $\sim20\,000$ stars like the real catalogue, were analysed in two ways: (i)
using all six phase-space coordinates $\vu$, and (ii) using only the five
astrometric coordinates to mimic the actual case in which spectra are not
available.

We used \textit{SLSQP} method of {\it scipy} \citep{Virtanen2020} to
locate a minimum of minus the log-likelihood, and started the MCMC search
from this DF. We searched with the {\it emcee} package \citep{Mackey2013},
with 35 walkers for each parameter and 500 steps in total. The first 50 steps
were considered burn-in and do not contribute to the statistics we present.

\subsection{Mock catalogue with full 6-D information}\label{sec:6d}

Fig.~\ref{fig:mock_6d} gives results from a single mock catalogue when using
the values for $v_\parallel$.  Each panel shows the marginalized posterior
probability distribution of a parameter. The parameter's true value is where
the blue lines intersect. The central dashed line in each one-dimensional
histogram marks the median of the distribution, and 68 percent of the
probability is contained between it and the dashed lines on either side,
at the
16th and 84th percentiles. These lines show that the MCMC exploration is
estimating the parameters within the expected uncertainties and with no
evidence of bias.  The second row of Table~\ref{tab:finalcom} summarises the
results shown in Fig.~\ref{fig:mock_6d}.

The only strong correlations are between the break action $J_0$ and the inner
and outer slope parameters $\alpha$ and $\beta$. The existence of this
correlation is entirely natural, and it gives rise to the weaker correlation
between $\alpha$ and $\beta$ shown in the top left 2-d histogram.

The coefficients $h_i$ and $g_i$ of the actions, which establish the velocity
ellipsoid, are recovered with good precision. The slope $\alpha$ of the inner
halo is also recovered well, while the slope $\beta$ of the outer halo, and
the break action $J_0$ are less precisely recovered. This is to be expected
because the stars in the outer halo have relatively large errors in velocity.

\subsection{Mock catalogue without line-of-sight velocities}\label{sec:5d}

Fig.~\ref{fig:mock_5d} is the analogue of
Fig.~\ref{fig:mock_6d}  for the case in which
line-of-sight velocities are not available, and the third row of
Table~\ref{tab:finalcom} gives the means and standard deviations of its
distributions.  We see that deleting the
line-of-sight velocities introduces no new correlations between the parameters
and has rather a small effect on the precision with
which the parameters of the DF can be recovered. One might say that the
line-of-sight velocities are almost redundant, a conclusion that parallels a
result of \cite{McMillan2012}.

\subsection{Sanity check}\label{sec:vali}

The results presented above involve the analysis of a single mock catalogue,
with and without the availability of line-of-sight velocities. In these
tests, both observational error and limited sample size contribute to the
breadth of the a posterior probability distributions shown in
Figs.~\ref{fig:mock_6d} and \ref{fig:mock_5d}, which in 
Table~\ref{tab:finalcom} are reduced to means and
standard deviations. However, the mean values of these probability
distributions are liable to be shifted one way or the other by the
peculiarities of the particular realisation of the DF that was analysed. We
now investigate the extent of such shifts by
examining the scatter in the mean values obtained from ten different
catalogues, each a different realisation of the same DF.

\begin{figure*}
	\includegraphics[scale=0.45]{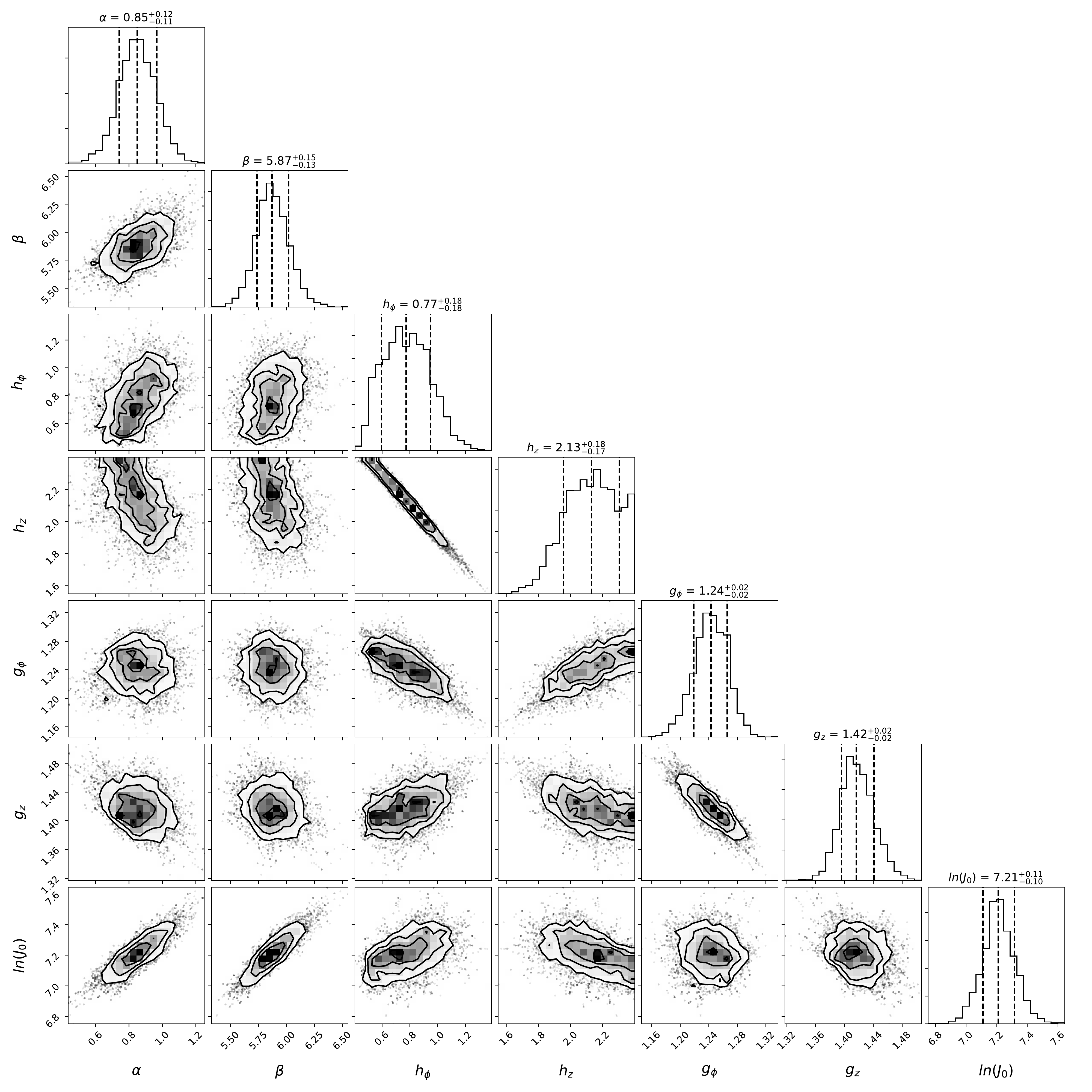}
    \caption{As Fig.~\ref{fig:mock_6d} but after fitting to the data for real
    rather than mock RR-Lyrae stars.}
    \label{fig:realresult}
\end{figure*}

\begin{table*}
	\centering
	\caption{Means and 1-$\sigma$ uncertainties of parameters fitted to
	the data for real RR-Lyrae stars.}
	\label{tab:realvalue}
	\resizebox{\textwidth}{!}	
	{\begin{tabular}{lccccccc} 
		\hline
		&$\alpha$ & $\beta$ & $h_{\phi}$  & $h_{z}$ &$g_{\phi}$
		& $g_{z}$ & $\ln{J_{0}}$\\
		\hline
		Fitted Values & $0.85^{+0.12}_{-0.11}$ &
		$5.87^{+0.15}_{-0.13}$ & $0.77^{+0.18}_{-0.18}$  
		& $2.13^{+0.18}_{-0.17}$ & $1.24^{+0.02}_{-0.02}$ 
		&  $1.42^{+0.02}_{-0.02}$ & $7.21^{+0.11}_{-0.10}$\\
		\hline
	\end{tabular}}
\end{table*}

The bottom two rows of Table~\ref{tab:finalcom} list from 6d and 5d analysis,
these means-of-means and standard deviations.  The standard deviations for
parameters other than $J_0$ and $\beta$ agree well with the 1-$\sigma$
uncertainties derived from a single catalogue.  The means of the correlated
parameters $J_0$ and $\beta$ both increase in line with the correlation found
between them, and the standard deviations of $\beta$, and to a lesser extent
$J_0$, exceed the 1-$\sigma$ uncertainties from a single catalogue.
Overall the ten-catalogue experiment confirms the reliability of statistics
extracted from a single catalogue.

\begin{figure}
	\centerline{\includegraphics[width=.8\hsize]{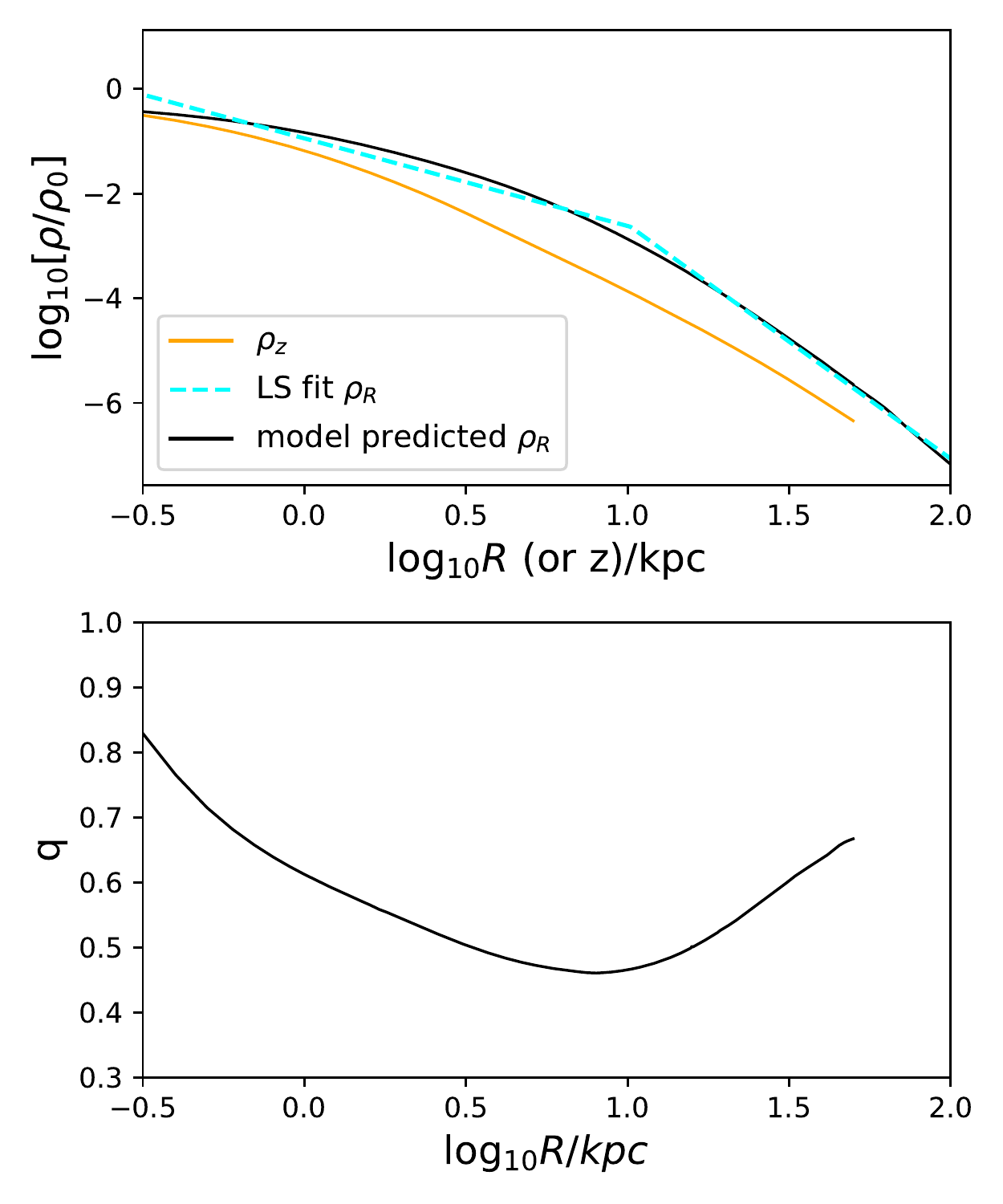}}
\caption{The density distribution predicted by the best-fitting DF. Top:
density along the major (black) and minor (orange) axes. The blue lines show
a double-power-law fit to the major-axis profile. Bottom: the resulting axis
ratio.}\label{fig:real_space}
\end{figure}

\begin{figure}
	\centerline{\includegraphics[width=.8\hsize]{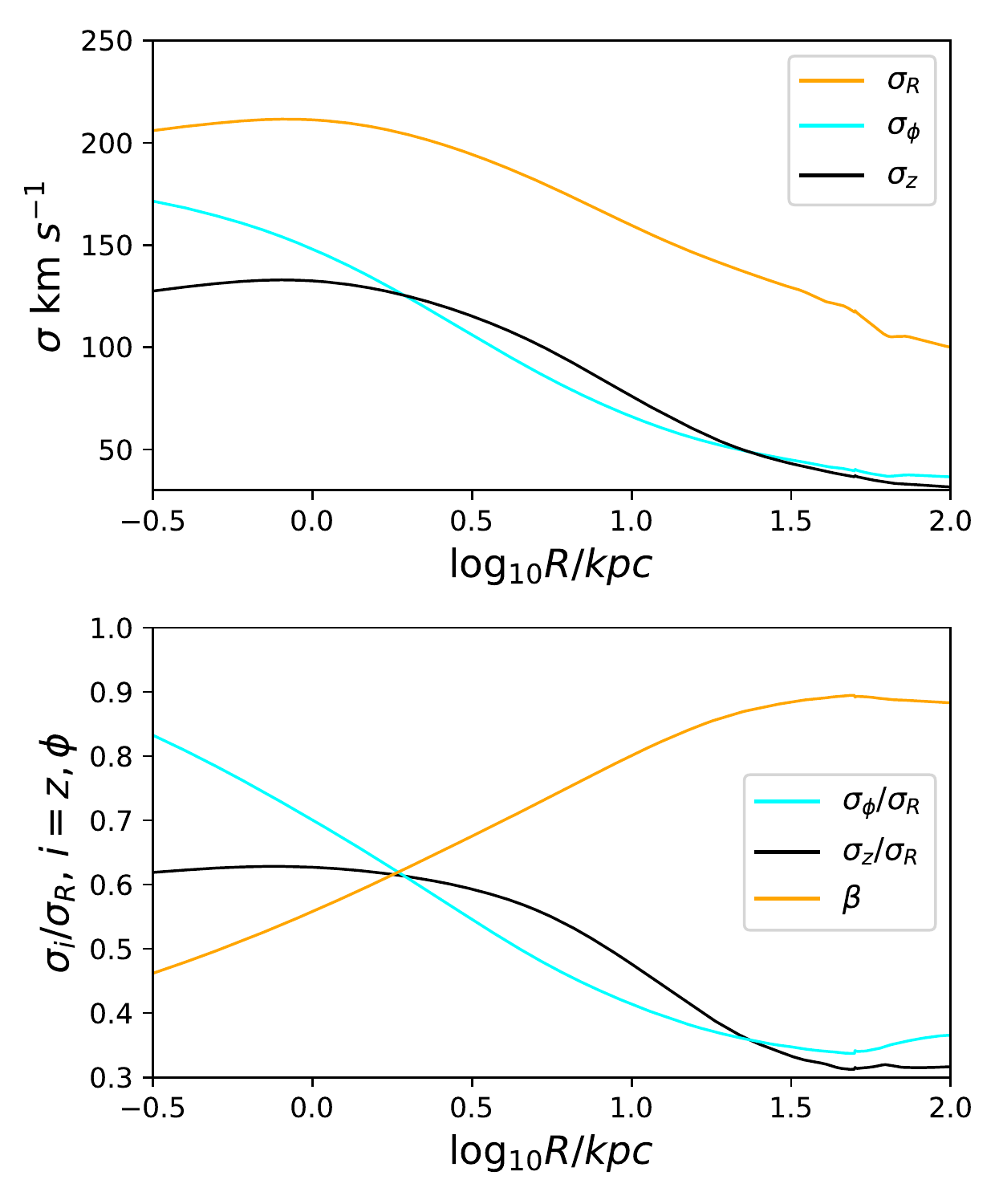}}
\caption{Velocity dispersions along the major axis of the best-fitting model.
The lower panel shows the ratios $\sigma_\phi/\sigma_R$ and
$\sigma_z/\sigma_R$ along the major axis.}\label{fig:velocity_space}
\end{figure}

\begin{figure*}
	\includegraphics[scale=0.65]{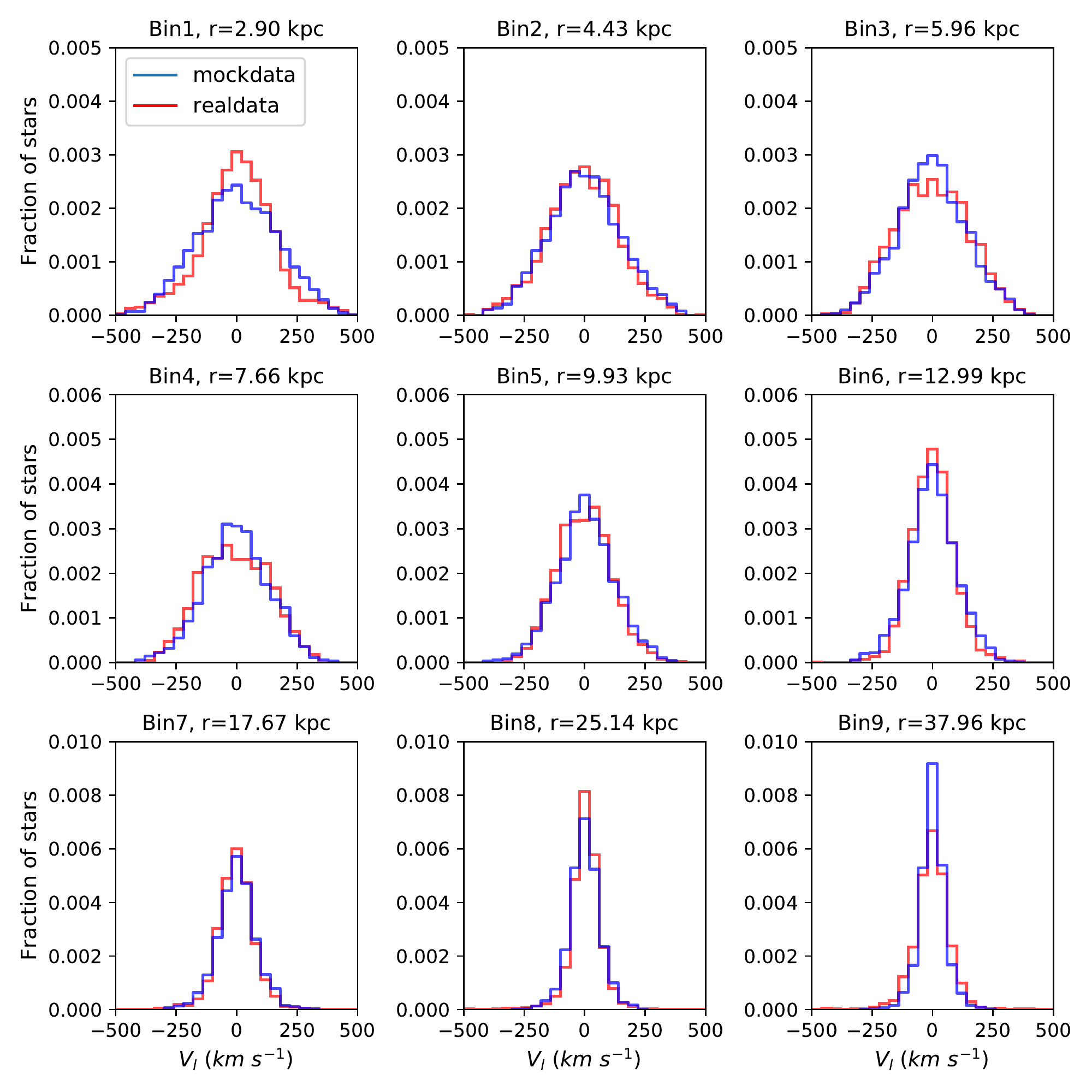}
    \caption{Distributions of $V_\ell$ for real RR-Lyrae stars (red histograms)
    and mock stars drawn from the best-fitting DF.
The radius marked in each bin is the median Galactocentric radius of the mock
stars in that bin.}
    \label{fig:vl}
\end{figure*}

\begin{figure*}
	\includegraphics[scale=0.65]{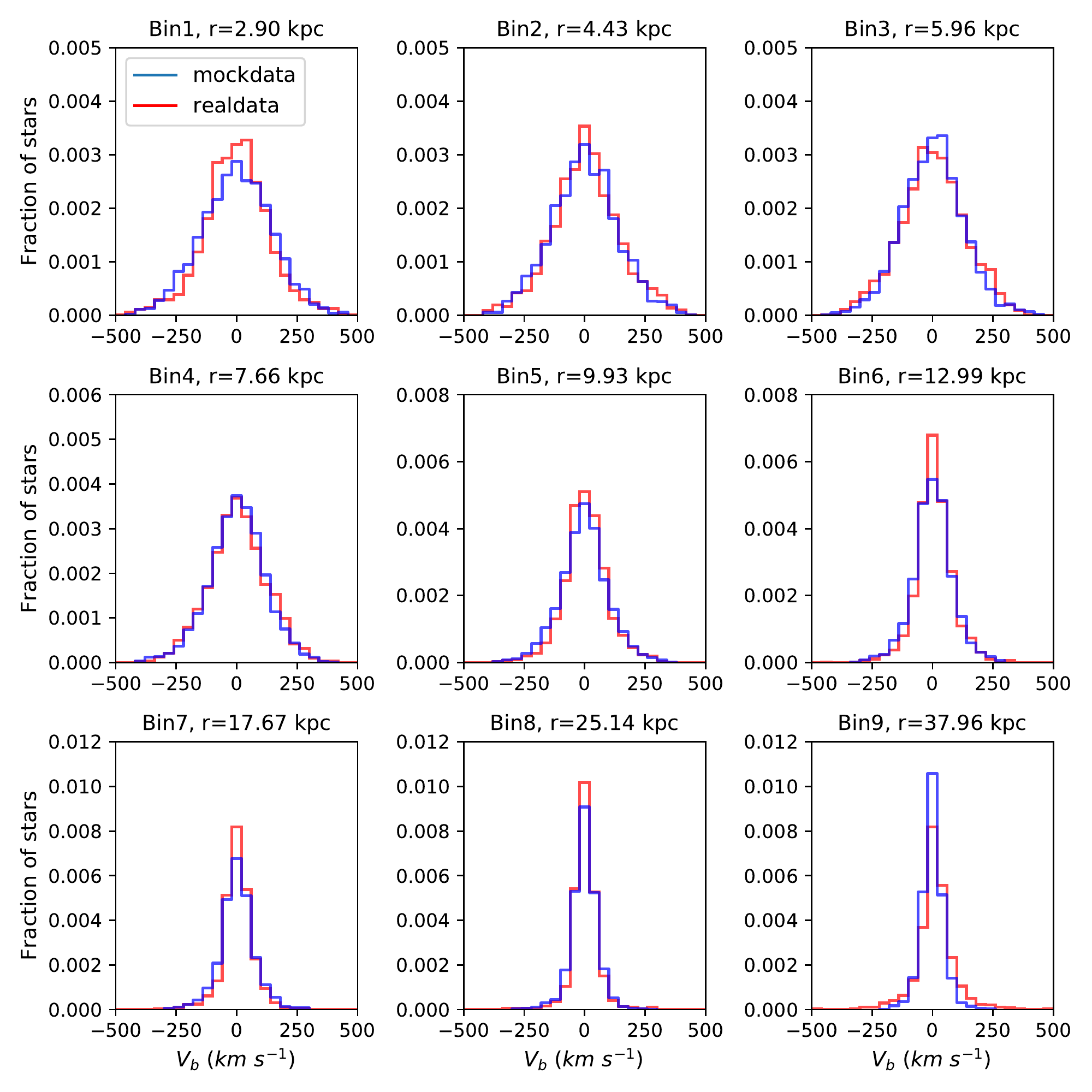}
    \caption{As Fig.~\ref{fig:vl} but for $V_b$.}
    \label{fig:vb}
\end{figure*}

\section{DF parameters from real RR-Lyrae stars} \label{sec:realdata}

We now apply the method for missing $v_\parallel$ to the real RRab catalogue from
PanSTARRS and Gaia EDR3. Fig.~\ref{fig:realresult} shows the distributions
and correlations of the parameters from the MCMC search.
Table~\ref{tab:realvalue} shows the means and standard deviations of these
distributions.

The principal difference between Fig.~\ref{fig:realresult} and the
corresponding figures from the tests is the appearance of a tight correlation
between the anisotropy parameters $h_\phi$ and $h_z$, which are subject to
the constraint $h_z+h_\phi<3$ so that $h$ is non-zero, and the DF is
finite, away from the origin of action space. In the tests, the underlying
DF is not strongly anisotropic and the constraint plays no role. The DF of
the RR-Lyrae stars is well known to be strongly radially biased, and the
constraint $h_\phi+h_z<3$ forces the probable values of the parameters to lie
close to the line $h_\phi+h_z=3$ along which the DF is almost independent of
$J_r$. Several other studies of halo populations using Gaia DR2 data have
concluded that the inner stellar halo is strongly radially biased
\citep{Belokurov2018,Wegg2019,Bird2019}. 

Table~\ref{tab:realvalue} indicates that for the RR-Lyrae stars
$h_{z}>h_\phi$, which implies that the RR-Lyrae population is flattened.
Several previous studies have reached this conclusion for sub-populations of
the stellar halo
\citep{Carollo2010,Sesar2011,Das2016a,Hernitschek2018,Wegg2019}.

The values $g_\phi=1.24$, $g_z=1.42$ and thus $g_r=0.34$ derived for
the outer halo indicate that the latter is also flattened but not so strongly
radially biased as the inner halo. Previous studies of the stellar halo have
concluded that its flattening and radial bias decrease with radius
\citep{Pila-Diez2015,Xue2015,Bird2019,Wegg2019,Bird2020}.

The uncertainties in the parameters that define the inner halo are larger
than we encountered with  the mock catalogues, while similar uncertainties
are found in the break action and the parameters that define the outer halo. 

\subsection{Spatial structure of the population}\label{sec:real_space}

The upper panel of Fig.~\ref{fig:real_space} shows the density profiles of the
RR-Lyrae population in the plane (black curve) and along the symmetry axis
(orange curve). The flattening of the model is evident from the fact that the
orange curve lies well below the black curve.  The cyan dashed lines in this
plot show a least-squares fit to the in-plane density profile with two power
laws:
\begin{equation}\label{eq:double_power_law}
    \rho(r)=\rho_0\times\begin{cases}
(r_{\rm b}/r)^{\alpha_{\rm i}}&r<r_{\rm b},\cr
(r_{\rm b}/r)^{\alpha_{\rm o}}&r>r_{\rm b}.
\end{cases}
\end{equation}
The break radius $r_{\rm b}$ is encoded in the DF by the break action $J_0$, while
the inner and outer density slopes $\alpha_{\rm in}$ and $\alpha_{\rm o}$ are
encoded in the DF by $\alpha$ and $\beta$. Their best-fit values are
$r_{\rm b}=10.2\kpc,\, \alpha_{\rm i}=1.7, \alpha_{\rm o}=4.5$.  A straight line
provides a good fit to the model density profile at $R>10\kpc$, but no
straight line could provide a good fit to the profile at smaller radii. The
significant curvature of the density profile inside the break radius reflects
the complexity of the potential in this region, where bulge, disc and dark
halo all contribute significantly -- \cite{Posti2015} showed that the DF we
are fitting to the RR-Lyrae stars generates a self-consistent model that is
well fitted by a double power law in radius, but our RR-Lyrae model sits in
an externally generated potential.

The break radius $r_{\rm b}\simeq10\kpc$ derived above is only half that
derived for the stellar halo in other studies
\citep{Sesar2011,Xue2015,Hernitschek2018}. We defer discussion of this issue
to Section~\ref{sec:prior}.

The outer slope in our model, $\alpha_{\rm o}=4.5$, is comparable to the
values estimated for other halo populations
\citep{Sesar2011,Xue2015,Hernitschek2018}. This parameter is vulnerable to
the use of an erroneous selection function.

The lower panel of Fig.~\ref{fig:real_space} shows the axis ratio $q$ of
isodensity surfaces as a function of semi-major axis length $R$. The
population is most flattened around the solar radius because it is there that
the disc's contribution to the overall potential is largest; at smaller and
larger radii the bulge and the dark halo dominate.

The upper panel of Fig.~\ref{fig:velocity_space} shows how the velocity
dispersions vary with radius $R$ along the major axis, and the lower panel
shows the axis ratios of the velocity ellipsoid. Even though $h_\phi+h_z$
lies near its upper bound of 3, while $g_\phi+g_z$ is significantly smaller,
the (strong) radial bias is outwards increasing. At $R\la1\kpc$,
$\sigma_\phi$ tends to $\sigma_R$, as it must in any plausible axisymmetric
model (Binney \& Vasiliev in preparation), but at all radii for which we have
observational data, both axis ratios are declining with increasing $R$ from
$\sim0.6$ to $\sim0.3$. These results are consistent with the findings of
\cite{Wegg2019}. In the lower panel of Fig.~\ref{fig:velocity_space},
the orange solid line shows the anisotropy parameter
$\beta=1-\fracj12(\sigma^2_\theta+\sigma^2_\phi)/\sigma^2_r$, which is as
large as $\sim0.9$ beyond $30\kpc$ from the Galactic centre. This is
consistent with recent studies of the stellar halo
\citep{Bird2019,LancasterKoposov2019,Bird2020}.

\section{Quality control}\label{sec:quality}

A best-fit model has significance only to the extent that it provides a
convincing fit to the data -- if a decent approximation to the true DF can be
obtained for no values of its parameters, the fitted DF is unlikely to be
able to predict the data even for the best-fit parameters. Hence we now 
ask whether the best-fit DF successfully accounts for the statistics of the
data by comparing the observational catalogue with a mock catalogue constructed
by sampling the best-fit DF.

Figs.~\ref{fig:vl} and \ref{fig:vb} compare real (red) and mock (blue)
distributions of $V_\ell$ and $V_b$, respectively. From top left to bottom
right the Galactocentric distance of the stars increases from $r>2.9\kpc$ to
$r<39\kpc$. The mock distributions were obtained in a two-step process.  
The DF was sampled to produce true values of the observables for the same
number of stars as there are in real catalogue.  The mock stars were then
binned in Galactocentric radius $r$ and the effect of observational errors
was simulated by smoothing as follows.

First the distributions of the uncertainties in $\mu_\alpha$ and $\mu_\delta$ for
stars in each bin were plotted and appropriate dispersions chosen. Then the
probability $P_{ij}$ that a star is scattered by an observational error
$\epsilon_j$ from the true value $x_j$ of an observable into a bin that has
boundaries $X_{i-1}<X_i$ is
\begin{equation}\label{eq:erroreachbin}
    P_i(x_j)=\int_{X_{i-1}}^{X_i} \frac{\d x}{\sqrt{2\pi}\epsilon_{j}}
    \e^{-{(x_j-x)^2}/{2\epsilon_j^2}}.
\end{equation}
If the true number density of stars is $F(x)$, then the number that will be
found in the $i$th bin is
\[
N_i=\int_{-\infty}^\infty\d x\,P_i(x)F(x)
\]
In the approximation that the $F(x)$ is constant across the width of each bin,
we have
\[
N_i\simeq\sum_j {n_j\over X_j-X_{j-1}}
\int_{X_{j-1}}^{X_j}\d x\,P_i(x)
\simeq\sum_j n_jP_i(\overline{x}_j),
\]
where $n_j=F(\overline{x}_j)(X_j-X_{j-1})$ with
$\overline{x}_j=\frac12(X_{j-1}+X_j)$ is the number expected in the
$j$th bin in the absence of errors. 

\begin{figure}
	\includegraphics[width=\columnwidth]{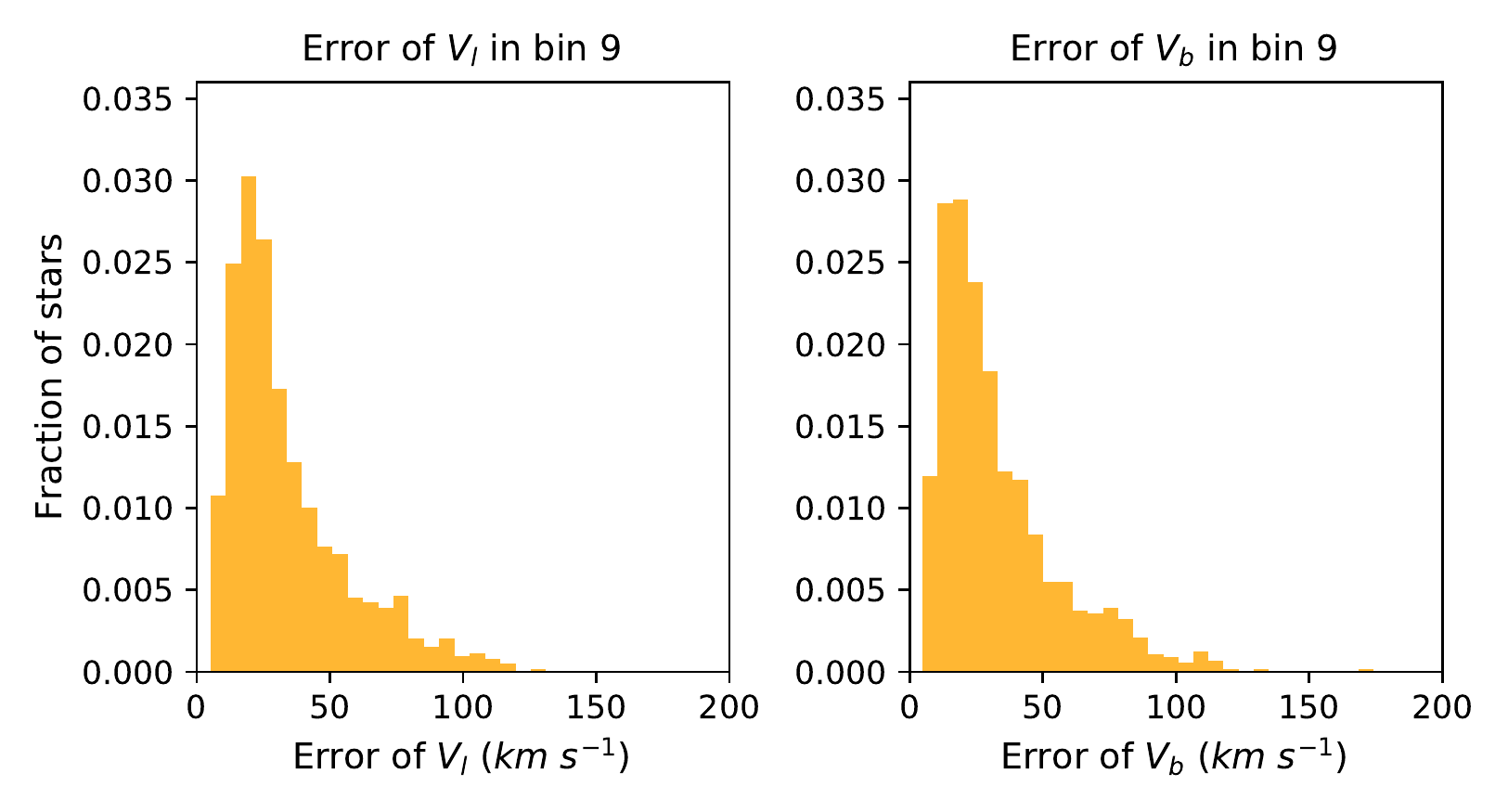}
    \caption{The errors of $V_\ell$ (left) and $V_b$ for real stars in the
    bin furthest from the Galactic centre.}
    \label{fig:errvlvb}
\end{figure}

\begin{figure}
	\includegraphics[width=\columnwidth]{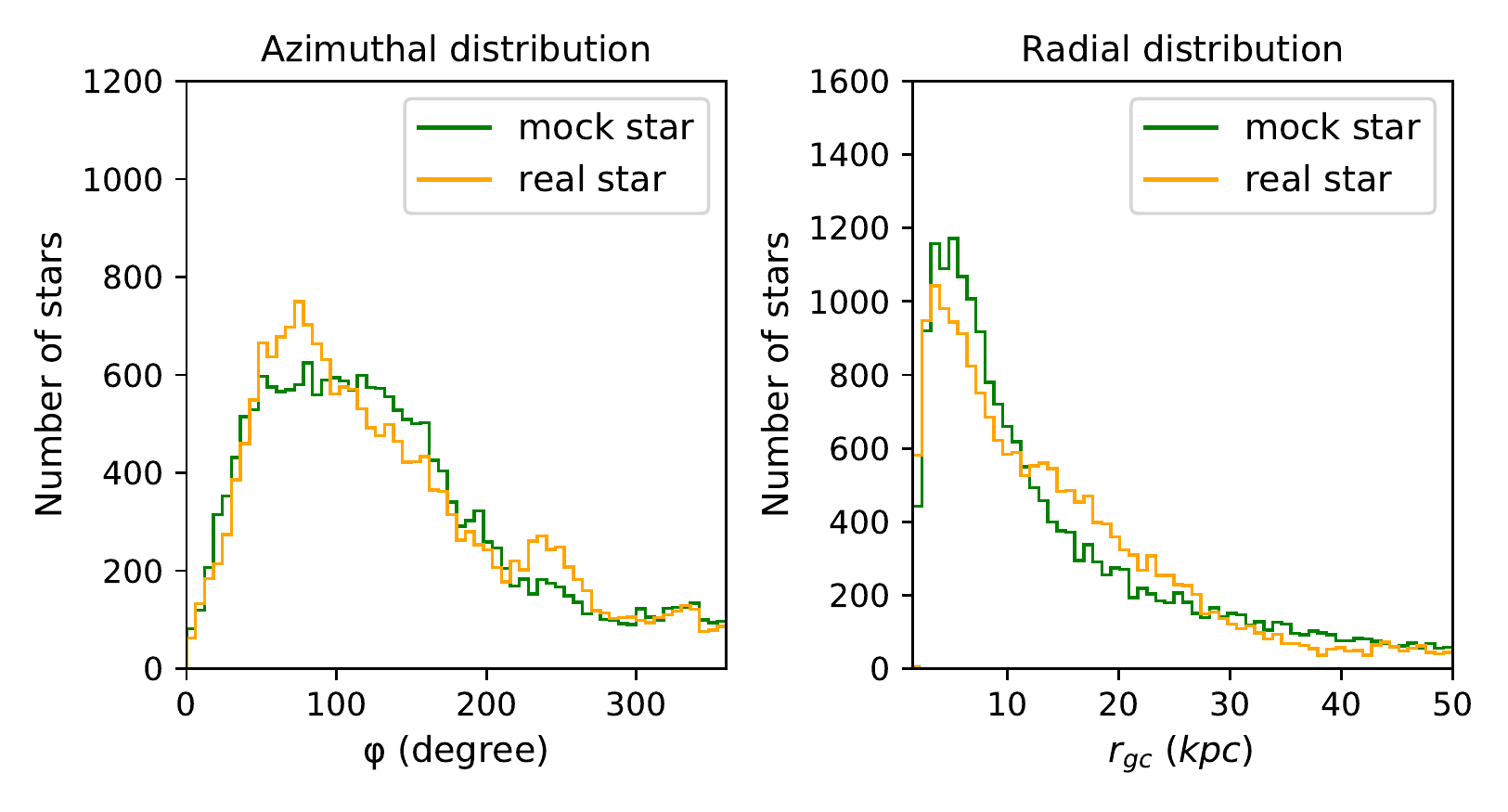}
	\includegraphics[width=\columnwidth]{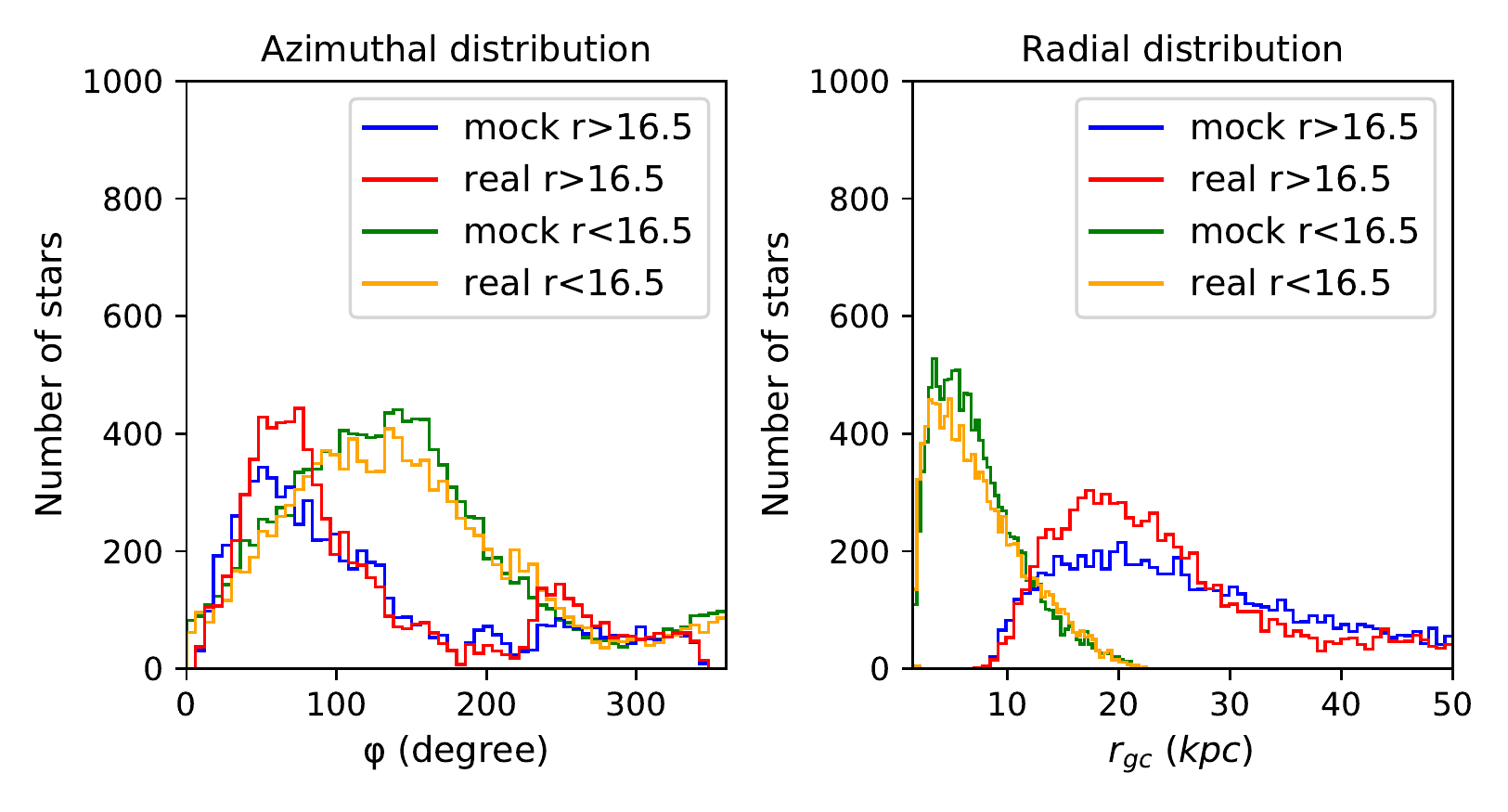}
    \caption{Distributions of  stars in Galactocentric azimuth
    $\phi$ (left column) and radius $r$ (right column). Top row: all real
    stars (orange) and mock stars (blue) when using the standard selection function. Bottom row: the same
    catalogues split into  brighter ($r<16.5$ orange/blue) and fainter
    (red/blue) stars.
}
    \label{fig:rphicom}
\end{figure}

\begin{figure*}
	\includegraphics[scale=0.65]{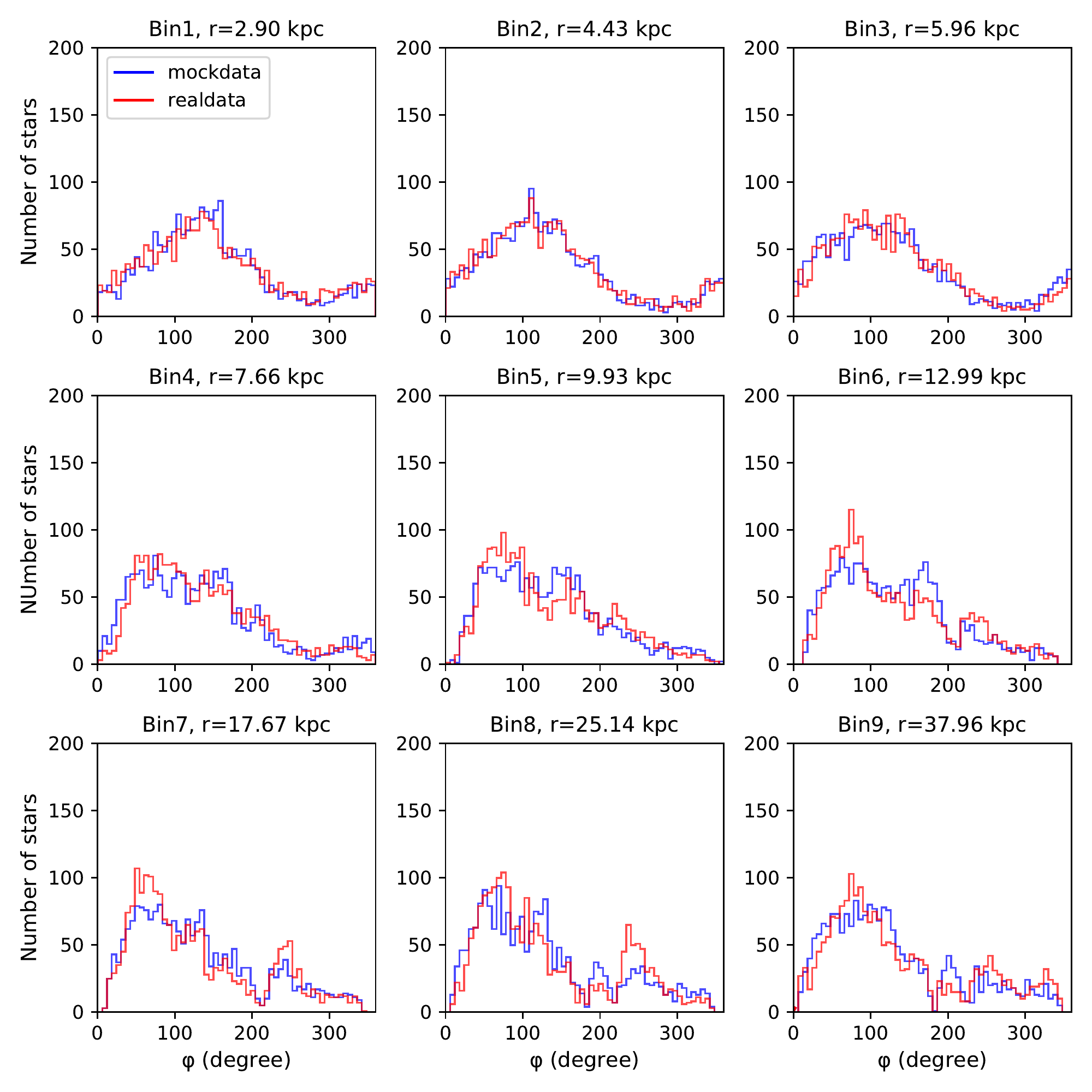}
    \caption{The distribution of $\phi$ for both real RR-Lyrae stars (red
    histogram) and  mock stars drawn from the best-fitting DF. The bins are
    the same as Figs.~\ref{fig:vl} and \ref{fig:vb}. }
    \label{fig:phi9}
\end{figure*}

In Figs.~\ref{fig:vl} and \ref{fig:vb} the observed (red) and predicted
(blue) histograms agree quite well at most radii. Both the observed and the
mock distributions become narrower as one proceeds to larger radii. This
trend reflects the marked falls in the velocity dispersions shown in the
upper panel of Fig.~\ref{fig:velocity_space}. With the possible exception of
the top left panel of Fig.~\ref{fig:vl}, the red histograms show no evidence
of being skew, as they would be if the RR-Lyrae population were
systematically rotating. In the top left panel of Fig.~\ref{fig:vl}, for
$r\sim2.9\kpc$, there is a hint of skewness, consistent with the finding of
\cite{Wegg2019} of rotation at $r<4\kpc$. Since we have chosen to use a DF
that is even in $J_\phi$, the blue histograms should be symmetric to within
Poisson noise. However, the skewness of the red histogram is not alone
responsible for the disagreement between the red and blue histograms in the
top left panel of Fig.~\ref{fig:vl}: the main problem is that the mock
distribution is too broad. The corresponding panel of Fig.~\ref{fig:vb} shows
a similar, but weaker, conflict. These conflicts may signal a problem with
the functional form of the DF at small $|\vJ|$. 

In Figs.~\ref{fig:vl} and \ref{fig:vb}, the bottom-right panels for the stars
furthest from the Galactic centre display the opposite effect: the red
histograms are now broader than the blue ones. From left to right along the
bottom rows of both figures, the blue histograms become steadily narrower in
line with the fall in $\sigma$ with increasing $r$. In the first two
panels, the red histograms follow suit, but in the third panels the red
histograms broaden. Is this phenomenon real or a reflection of problems with
the data? In the former case, we should modify the DF at large $|\vJ|$.
On the other hand, the data for these most distant stars $\ex{r}\simeq39\kpc$
are subject to large uncertainties, as is illustrated by the broad and highly
non-Gaussian distribution of uncertainties in $V_\ell$ and $V_b$ plotted in
Fig.~\ref{fig:errvlvb}. Moreover, stars with the most over-estimated distances
will accumulate in this bin alongside inflated values of $V_\ell$ and $V_b$.

The upper row of Fig.~\ref{fig:rphicom} compares the observed (orange) and
mock (blue) distributions in Galactocentric azimuth $\phi$ (left panel) and
radius $r$.  Systematic differences in the distributions are evident. The
lower row of Fig.~\ref{fig:rphicom} probes the origin of these differences by
showing the azimuthal and radial distributions separately for stars
brighter/fainter than magnitude $r=16.5$. We see that the mock and observed
histograms for the brighter stars agree well and the discrepancies lie with
the distributions of faint stars: in the lower right panel of
Fig.~\ref{fig:rphicom}, the mock sample
(blue) contains too few faint stars at $r_{\rm gc}\sim20\kpc$ and a slight
excess at large radii. This issue cannot be resolved by simply changing the
model's density profile because the yellow and green curves for bright stars
agree well at $r\sim20\kpc$: at this radius the model works well in the
anticentre direction but at most distant locations it predicts too few stars.
The lower left panel shows that the deficiency is concentrated at longitudes
$\ell=80$ and 240 degrees.  Two explanations are in principal viable: the
stellar halo, unlike the model, is not axisymmetric, having enhanced density
at $\ell\sim80$ and 260 degrees, or the (very much non-axisymmetric)
selection function that we have applied to the model is incorrect and culls
too many stars at the given longitudes.




\subsection{Overdensities}

Fig.~\ref{fig:phi9} compares the observed and mock distributions in
Galactocentric azimuth $\phi$ when the sample is divided into bins by
Galactocentric radius as in Figs.~\ref{fig:vl} and \ref{fig:vb}.  The
agreement is particularly good out to the solar radius, but from the fifth
bin ($r\sim10\kpc$) onwards the red histograms for real stars show more
structure than the mock blue histograms. Is this additional structure caused
by stellar streams that we have not eliminated from the data? Smooth
structure at $r<R_0$ and sharp local peaks in density further out is exactly
the pattern that streams would be expected to generate.

It is interesting that the blue histograms do have local peaks that
correspond to peaks in the red histograms (e.g., at $\phi\sim180^\circ$ in
the central panel and at $\phi\sim230^\circ$ in the leftmost panels of the
bottom row). The conflict between the histograms is just that these features
have larger amplitude in the red than in the blue histograms. Streams
contribute to the blue histograms in that our selection function, which culls
mock stars sampled from the DF, is designed to exclude streams. Since our
selection function includes some but not all streams, it is to be expected
that the mock and true histograms show similar structure, with 
more structure in the true histograms.

Streams other than Sgr were cut from the data using tables
in \citet{Yang2019}, which were compiled from LAMOST spectroscopic data.
LAMOST operates at rather bright magnitudes and has a more complex selection
function than the photometric surveys PanSTARRS and Gaia.
Consequently, our data probably do contains streams that were missed by
\citet{Yang2019}, particularly at faint magnitudes $r\sim18.5$.

\begin{figure}
\includegraphics[width=\columnwidth]{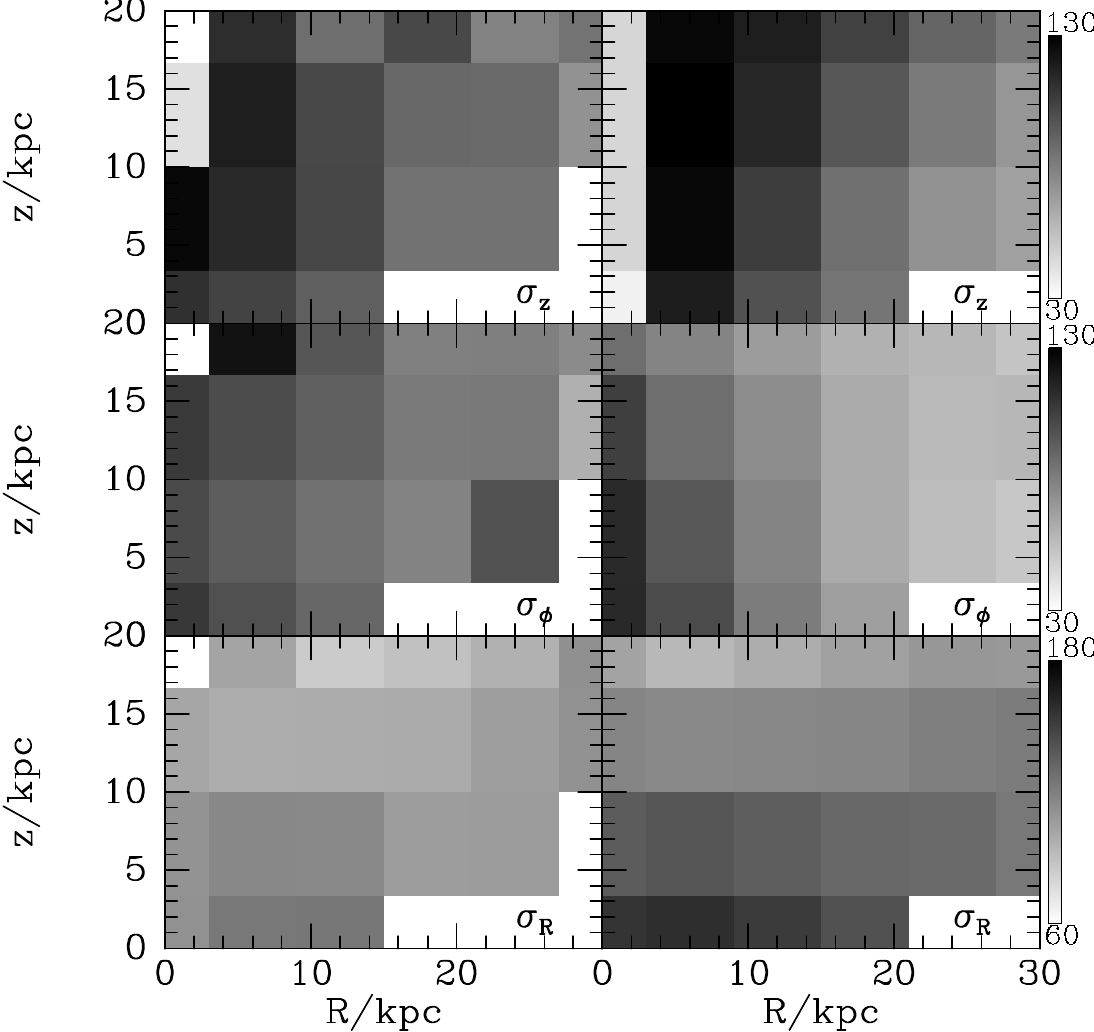}
    \caption{Left column: the principal velocity dispersions (in spherical
    coordinates) of the observed
    BHB stars when binned in $(R,|z|)$. Right column: the dispersions predicted
    by our model of the RR-Lyrae population.}
    \label{fig:testbhb}
\end{figure}

\begin{figure*}
\includegraphics[width=.7\hsize]{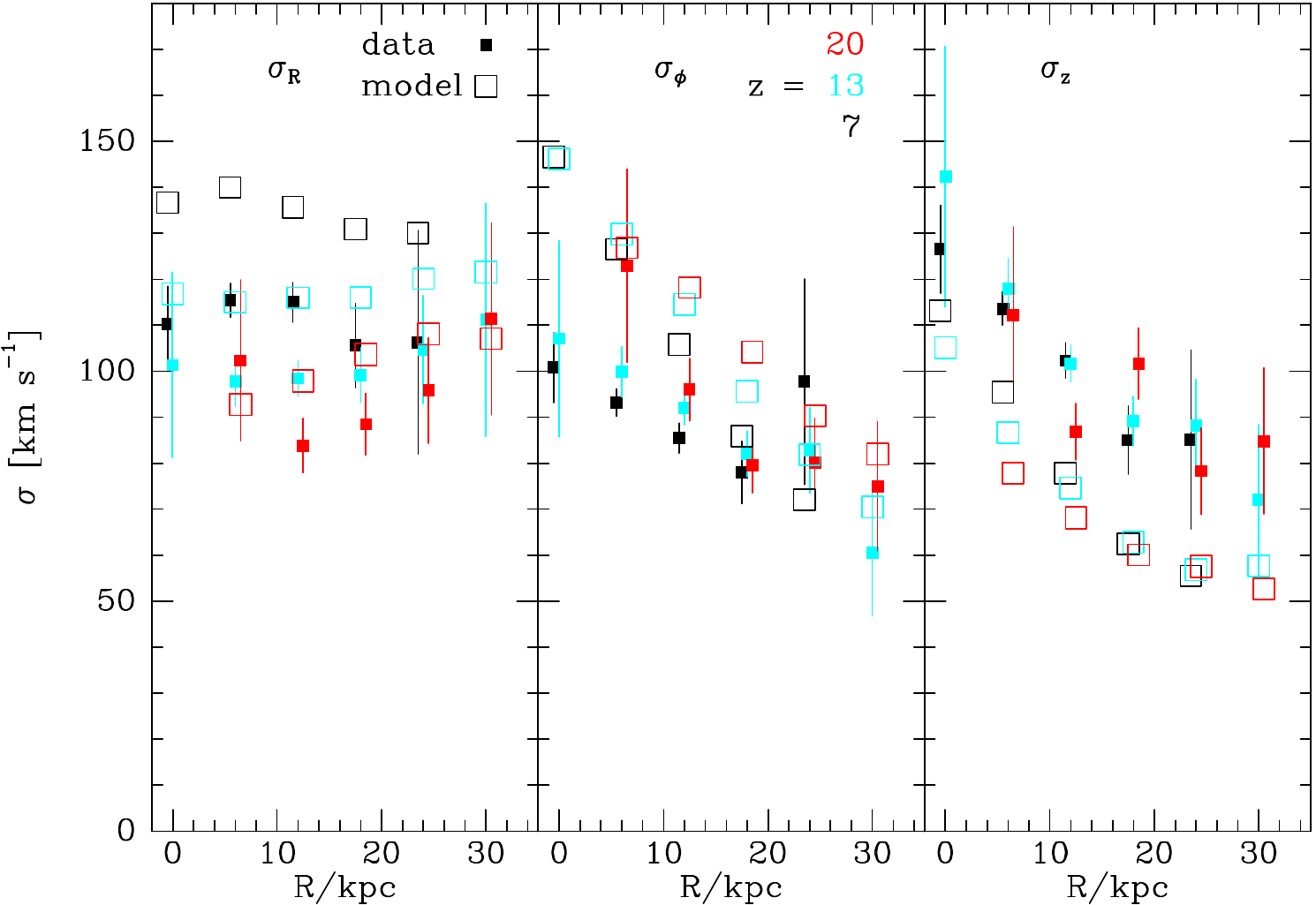}
\caption{Velocity dispersions of the BHB population along three
lines of constant distance from the plane, as functions of cylindrical radius
$R$. The bins are those shown in Fig.~\ref{fig:testbhb} less the bottom row
in each panel. Error bars are only shown for the data -- there are 20 times
as many mock stars as real ones, so the uncertainties of the model dispersions
are smaller by a factor $\sim4.5$.}\label{fig:testbhb2}
\end{figure*}

\section{Does the model fit BHB stars?}\label{sec:bhbtest}

RR-Lyrae stars might be expected to trace the metal-poor halo, and in this
section we compare our model of the RR-Lyrae population with the sample of
nearly 5000 BHB stars from \citet{Xue2011}. Accurate spectrophotometric
distances and line-of-sight velocities are available for BHB stars, and we
complement these with proper motions from Gaia EDR3. We exclude stars that
appear unbound or lie close to a globular cluster or within the Sagittarius
stream. We further exclude stars with very uncertain velocities by imposing
the criterion
\[
\sqrt{\epsilon^2_{v_x}+\epsilon^2_{v_y}+\epsilon^2_{v_z}}<50\kms.
\]
After these cuts, 2231 BHB stars remain in the sample.

The selection function involved in the identification of BHB stars by
\citet{Xue2011} is complex and poorly determined, so we will not concern
ourselves with the spatial distribution of the BHB stars. Fortunately, the
selection function should be blind as regards velocity, so we ask to what extent our
model predicts the velocity distribution of the BHB stars.

At the location of each BHB star, we instructed {\it AGAMA} to sample
the velocity distribution in our model and thus produce 20 mock BHB stars
at that location. Then we binned both the real and mock BHB stars in a grid in
the $Rz$-plane and compared the $Rz\phi$  velocity dispersions of the two
samples in cells that contain a useful number of real BHB stars.

Figs~\ref{fig:testbhb} and \ref{fig:testbhb2} show the results. The top row
in Fig.~\ref{fig:testbhb} and the extreme right panel of
Fig.~\ref{fig:testbhb2} show that the model predicts $\sigma_z$ quite well.
The middle panels of both figures indicate that at Galactocentric radii
similar to $R_0$ the model values for $\sigma_\phi$ are on the high side,
while the bottom panels of Fig.~\ref{fig:testbhb} and the extreme left panel
of Fig.~\ref{fig:testbhb2} indicate that near the plane the model values of
$\sigma_R$ are too large.

The stellar halo is now recognised to contain sub-populations. In particular,
recent work has identified a sub-population, the `Gaia sausage' of relatively
metal-rich stars on highly eccentric orbits that is speculated to have formed
during an early major merger
\citep{Helmi2018,Belokurov2018,LancasterKoposov2019,Iorio2021,Wu2021arxiv}.
In the likely event that the proportions of sausage stars in the RR-Lyrae and
BHB populations are different, one would expect RR-Lyrae and BHB kinematics
to differ. Hence the differences between model and data in
Figs.~\ref{fig:testbhb} and \ref{fig:testbhb2} do not necessarily reflect
weaknesses of the model. When spectra are available for a significant
fraction of the RR-lyrae stars, it is likely that a more complex model of the
RR-Lyrae population will be required in chemically identified sub-populations
have distinct kinematics.

\section{Prior work}\label{sec:prior}

\subsection{Wegg et al.\ 2019}

\citet[][hereafter W19]{Wegg2019} modelled a very similar data set for
$15\,813$ RR-Lyrae stars at $r<20\kpc$ in a different way: they fitted a
parametrised model
\[\label{eq:Wegg_profile}
\nu(r,\theta)=r^{-\alpha}\Big(\sin^2\theta+{\cos^2\theta\over
q^2}\Big)^{-\alpha/2}
\]
to the stars' spatial
distribution and assumed that at each location the density of stars in
velocity space is a function $F(Q)$ of
\[
Q(\vv)\equiv\fracj12(\vv-\overline{\vv})^T\cdot\vK\cdot(\vv-\overline{\vv}),
\]
where $\overline\vv=v_\phi\ve_\phi$ is the mean azimuthal streaming velocity
and 
\[
\vK\equiv\left(\begin{matrix}\sigma_{rr}^{-2}&\sigma_{r\theta}^{-2}&0\cr
\sigma_{r\theta}^{-2}&\sigma_{\theta\theta}^{-2}&0\cr
0&0&\sigma_{\phi\phi}^{-2}
\end{matrix}\right).
\]
The function $F$, $\vv_\phi$ and the four independent elements of $\vK$ were
determined by distributing the stars over $n_r\times n_\theta$ bins in the
meridional plane. This procedure has the merit that it does not require prior
specification of the Galaxy's potential.  Indeed, W19 were able to constrain
the Galaxy's gravitational field by applying the Jeans equations to an
estimate of the stellar stress field $(\nu\vsigma^2)(\vx)$ that they obtained
from equation \eqrf{eq:Wegg_profile} and their grid of values for
$\vsigma^2$. 

W19 adopted $n_r=8$ and $n_\theta=5$ and hence invoked $2+5n_rn_\theta=202$
free parameters. It is not a given that the density in velocity space is a
function of the quadratic form $Q$ and the imbalance in the distribution of
parameters between structure in real space (2 parameters) and velocity space
(200 parameters) is surely not ideal. Clearly W19 could not show a figure
analogous to Figs.~\ref{fig:mock_6d}, \ref{fig:realresult} and
\ref{fig:mock_5d} for 202 parameters, but they do say that many of their
parameters are highly correlated. 

In formulating a model, the goal should be to obtain a good fit to the data
using the smallest number of least correlated parameters. From this
perspective our model with seven parameters and only one significant
correlation in the tests is at an advantage to that of W19. Unfortunately,
when we applied the model to the real data a second, even stronger
correlation appeared, that between $h_\phi$ and $h_z$. This correlation
points to a fundamental weakness in the procedure for introducing velocity
anisotropy that was proposed by \cite{Binney2014}. This issue will be
addressed in forthcoming paper. Moreover, our DF is an even function of
$J_\phi$ and thus excludes the possibility of systematic rotation, for which W19
found evidence at $r\la4\kpc$. In future work we will increase our parameter count by
adding a part odd in $J_\phi$ to the DF. Even after a modest increase in parameter
count, the present method will  still involve an order-of-magnitude fewer
parameters  that that of W19.

\subsection{Pros and cons of modelling the DF}

The $f(\vJ)$ modelling technique holds down the number of parameters required
to fit rich data by exploiting Jeans' theorem, which connects real space to
velocity space. The price one pays for this advantage is that to predict
observables one requires both a DF and a potential $\Phi(\vx)$. When
modelling the entire Galaxy, $\Phi$ can be derived from the DF by the
self-consistency principle \citep{Binney2014,Binney2015,ColeBinney}. When
using a tracer population, the $f(\vJ)$ approach requires one to investigate
how well one's form for $f(\vJ)$ can fit the data when different potentials
are used. An exploration of this process using mock data by
\cite{McMillan2013} showed that care is needed to keep Poisson noise under
control. We hope shortly to constrain our Galaxy's potential in this way
using the present RR-Lyrae data and improved forms of the DF. We expect to be
able to break the existing degeneracy between the dark halo and the disc in
which a more flattened dark halo and less massive disc generates similar
near-plane kinematics to a less flattened dark halo and more massive disc
\citep{Binney2015}.

The real-space approach of W19 allows for a more direct assault on the constraint of
$\Phi$ but it is still fraught with difficulty because the Jeans equations
require derivatives of the stress $\nu\vsigma^2$, which can only be obtained
by binning stars on a grid.  Refining the grid increases both the parameter
count and the amplitude of Poisson noise. When a coarse grid is used the
inferred potential depends significantly on the interpolation scheme used to
extract derivatives.  Moreover, the pressure gradient that betrays the
gravitational field depends more strongly on the density $\nu$ than on the
velocity dispersion tensor $\vsigma^2$, so assigning two parameters to the
former and 200 parameters to the latter cannot be optimal.

As the density and flattening profiles of our model plotted in
Fig.~\ref{fig:real_space} illustrate, a DF $f(\vJ)$ which is a simple function
of the actions can generate a complex three-dimensional structure because the
latter reflects the structure of the potential in addition to that of the DF.
Whether this feature of $f(\vJ)$ modelling is a strength or a weakness hinges
on whether real components have simpler DFs than density profiles. We suspect
that they do, but this view is open to challenge.

\subsection{Break radius}

The density profile \eqrf{eq:Wegg_profile} adopted by W19 is a single power
law, and they derived slope $\alpha=2.6$. Similar results with stellar
haloes modelled by a single power law $\rho\propto r^{-\alpha}$ were obtained
by \citet{Iorio2018} and \citet{Mateu2018}, who found  $\alpha=2.96$
and 2.78, respectively. A simple power inevitably predicts divergent mass as
either $r\to0$ or $r\to\infty$ (or both when $\alpha=3$).  Hence most authors
have fitted broken power laws to components of the stellar halo. Our break
radius $r_{\rm b}\sim10\kpc$ is significantly smaller than previously
reported values.

\cite{Hernitschek2018} derived a break radius $38.7\kpc$ but this is the
radius of a transition from a steeper slope ($\alpha=4.97$) at $r<r_{\rm b}$ to a
shallower slope ($\alpha=3.93$) at $r>r_{\rm b}$. Clearly, the steep slope
$\alpha=4.97$ cannot continue to the origin. We do not feel able to
constrain the density profile at $r>39\kpc$, and in the range $10\la r/\!\kpc
\la 40$ our slope $4.5$ is broadly consistent with the result of
\cite{Hernitschek2018}.

Several recent studies derive radii at which the slope of a halo
population steepens outwards. \citet{Watkins2009} fitted a double power-law to
RR-Lyrae stars in SDSS, and found that at $r_{\rm b}=25\kpc$ the index
steepens from $2.5$ to $4.5$. \cite{Sesar2011} give a break radius $r_{\rm
b}=27.8\kpc$ at which the slope of the RR-Lyrae population increases from
$\alpha=2.62$ to $3.8$. \citet{Sesar2013} found that at $r_{\rm
b}=18\kpc$ the logarithmic slope steepens from $1.6$ to $3.4$. From a
sample of K giants \cite{Xue2015} found that at $r_{\rm b}=18\kpc$ $\alpha$
increases from $2.1$ to $3.8$.

\begin{figure}
\includegraphics[width=\columnwidth]{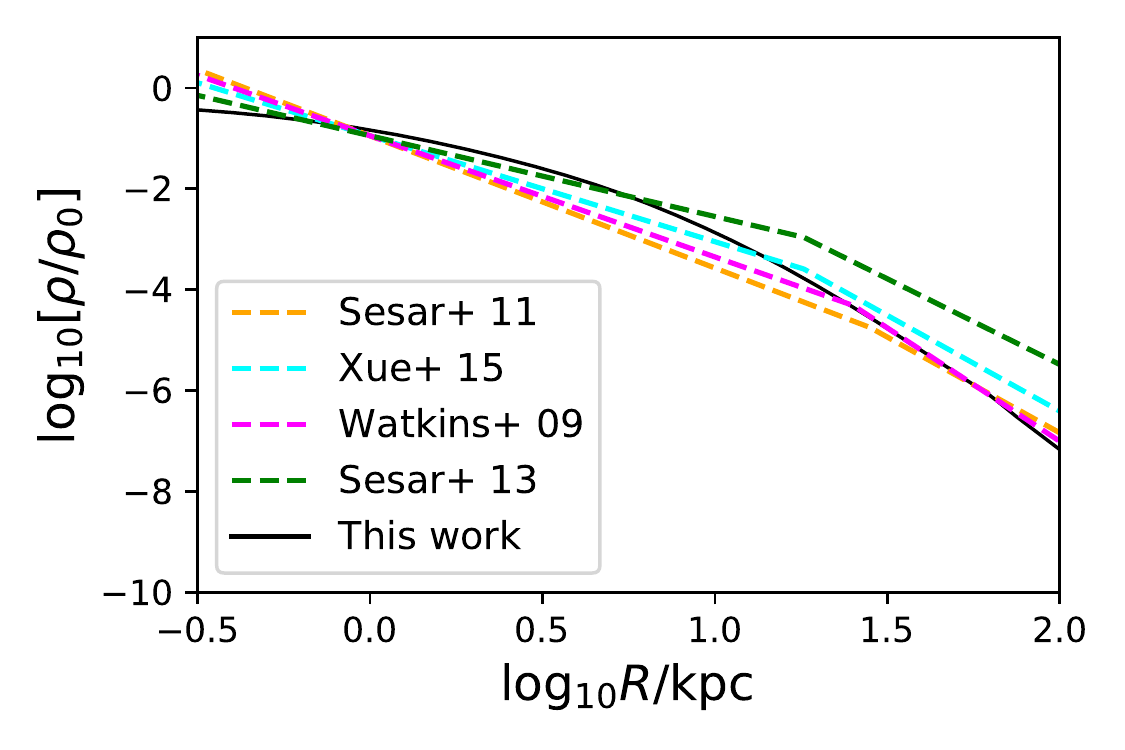}
    \caption{Comparisons between our best-fit model and four prior works \citet{Watkins2009}, \citet{Sesar2011}, \citet{Sesar2013}, and \citet{Xue2015}.}
    \label{fig:rhocompare}
\end{figure}

Fig.~\ref{fig:rhocompare} compares our density profile
with those of  \citet{Watkins2009}, \cite{Sesar2011},
\citet{Sesar2013}, and \cite{Xue2015}.
The densities predicted
by these curves disagree with one another only slightly less than they do
with the densities predicted by our model, which are shown by the full black
curve. Thus break radii that differ by nearly a factor 3 do not necessarily
imply significantly differing observationally testable predictions.

\section{Conclusions}\label{sec:conclusion}

We have fitted an $f(\vJ)$ model to the Galaxy's population of RR-Lyrae stars
using a sample of $\sim20\,000$ stars for which precise distances are
available but line-of-sight velocities are lacking. This exercise is of
interest both for the light it casts on the structure of our Galaxy's stellar
halo and as an exploration of what can be achieved in the absence of
line-of-sight velocities. Learning to exploit such data is important given
that line-of-sight velocities are unlikely ever to be available for the vast
majority of the $\sim1.3$ billion stars for which Gaia is harvesting
astrometry.  \cite{McMillan2012} suggested that one could deal with missing
data simply by setting to infinity the error on quantities not measured.
While correct in principle, this idea does not work in practice.  The
procedure developed here for dealing with missing data may prove the most
valuable contribution of this paper.

We fit a DF to the data by MCMC exploration of the seven-dimensional space of
model parameters. We have tested our code's ability to infer the parameters
of the model with which pseudo data was generated. We did this both for the
case in which line-of-sight velocities are available and when they are not.
In every case the true parameters lay within the probable region of parameter
space identified by the code, and deleting the line-of-sight velocity
information enlarged the probable region to a surprisingly small extent. Thus
when accurate astrometry is available, line-of-sight velocities are a luxury
without which it is straightforward to construct a population's DF.

In the tests, the only parameters that are significantly correlated are the
slopes of the DF at small and large $|\vJ|$, and the break action $J_0$ that
divides their two regimes. A correlation of this type is invariably
encountered when fitting a double power-law model to data. With the given
sample size and observational errors characteristic of Gaia EDR3, the only
parameter that is not tightly constrained is the break action
$J_0$, which has a 1-$\sigma$ uncertainty $\sim10$ per cent.

The standard Bayesian technique used here yields posterior probability
distributions of parameters that are broadened by both experimental error and
Poisson noise.  Hence, the means and standard deviations of parameters
extracted from a single realisation should agree with the standard deviation
obtained from the means of a sequence of different realisations of different
models. Table~\ref{tab:finalcom} tests this conjecture and confirms it with the
possible exception of the correlated parameters $\beta$ and $J_0$.

When we apply the code to the real sample of RR-Lyrae stars, there is one
material difference: now the parameters $h_\phi$ and $h_z$ that characterise
the velocity ellipsoids and flattening of the inner halo become strongly
anti-correlated. In fact the data push them up against the limit
$h_\phi+h_z<3$ that one imposes to ensure that the DF will not diverge at
finite $|\vJ|$. This result reflects the strong radial bias of the inner
halo, a phenomenon that has been interpreted as due to an early merger with the
`Enceladus' galaxy \citep{Belokurov2018}. The appearance of a strong
correlation between parameters signals the need for improved functional forms for
$f(\vJ)$ when fitting strongly anisotropic components. This need will be
addressed shortly.

Our model of the RR-Lyrae population is a flattened spheroid in which the
density settles to $\rho\sim r^{-4.5}$ at radii $r\ga10\kpc$. At smaller
radii the density is not well fitted by a power law although at $r<0.3\kpc$,
$\d\ln\rho/\d\ln r\sim-1.7$. In the equatorial plane
$\sigma_\theta\ga0.4\sigma_r$, so the model is not exceptionally strongly
radially biased, notwithstanding the limiting value of $h_\phi+h_z$. In the
region probed by the data, the ratios $\sigma_\phi/\sigma_r$ and
$\sigma_\theta/\sigma_r$ fall from $\sim0.6$ to $\sim0.3$ consistent with the
results of \cite{Wegg2019}.

A key issue when fitting a model to high-dimensional data by likelihood
maximisation is whether the best-fitting model provides an acceptable fit to
the data. Our model provides adequate fits to histograms of the tangential
velocities of stars in bins of Galactocentric radius except for stars nearest
and furthest from the Galactic centre. These discrepancies may may reflect
the very uncertain velocities of these stars. The distributions in
both radius and azimuth of bright stars are accurately predicted by the
model, but the model predicts too few faint stars at radii $r\sim20\kpc$ at
longitudes $\ell\sim80$ and 240 degrees. Merely changing the model's radial
density profile cannot resolve this issue; the problem could lie with our
selection function, or reflect departure of
the stellar halo from axisymmetry.

Likely footprints of stellar streams are visible in plots of the azimuthal
distributions of stars binned by Galactocentric radius $r$. Inside the solar
sphere we don't expect significant contributions from streams, and indeed at
these radii the mock and real catalogues agree nicely. At larger radii the
real histograms show more substructure than the mock ones, as would be
expected if streams contribute to the data despite our efforts to exclude
known streams.  The mock histograms have less pronounced local maxima at
similar azimuths to the real histograms, perhaps because the mock data were
filtered by a selection function designed to mask streams.

We used our model of the RR-Lyrae population to predict the velocity
distribution at the locations of BHB stars with measured space velocities.
 The $z$ velocities are predicted well but too much elongation was
predicted in the $v_R v_\phi$ plane.  This problem may reflect real
differences between the BHB population and the slightly more metal-rich
RR-Lyrae population, or a poor choice for the form of the DF.

This work constitutes a proof of principle: five-dimensional data such as
Gaia provides for over a billion stars can be used to build a dynamical model
of a stellar component. This opens up two avenues for further work. First
improved functional forms of the DF will be fitted to the present sample of
RR-Lyrae stars in the potential generated by a series of self-consistent
Galaxy models. A prime focus of this  work will be on constraining the shape
and local density of the dark halo by using the RR-Lyrae stars to probe the
potential further from the plane and out to larger radii than stars in Gaia's
RVS sample do.

A second natural application of the approach we have developed here would be
use stars not in the RVS catalogue alongside RVS stars in the construction of
self-consistent Galaxy models.

\section*{Acknowledgements}
We thank the anonymous referee and the scientific editor for the suggestions to improve this draft. We thank D.\ Boubert for pointing out the possibility of refining proper
motions when an accurate distance is available. 
CL and JB are supported by the UK Science
and Technology Facilities Council under grant number ST/N000919/1. JB also
acknowledges support from the
Leverhulme Trust through an Emeritus Fellowship. 

This work presents results from the European Space Agency (ESA) space mission Gaia. Gaia data are being processed by the Gaia Data Processing and Analysis Consortium (DPAC). Funding for the DPAC is provided by national institutions, in particular the institutions participating in the Gaia MultiLateral Agreement (MLA). The Gaia mission website is https://www.cosmos.esa.int/gaia. The Gaia archive website is https://archives.esac.esa.int/gaia.

The Pan-STARRS1 Surveys (PS1) and the PS1 public science archive have been made possible through contributions by the Institute for Astronomy, the University of Hawaii, the Pan-STARRS Project Office, the Max-Planck Society and its participating institutes, the Max Planck Institute for Astronomy, Heidelberg and the Max Planck Institute for Extraterrestrial Physics, Garching, The Johns Hopkins University, Durham University, the University of Edinburgh, the Queen's University Belfast, the Harvard-Smithsonian Center for Astrophysics, the Las Cumbres Observatory Global Telescope Network Incorporated, the National Central University of Taiwan, the Space Telescope Science Institute, the National Aeronautics and Space Administration under Grant No. NNX08AR22G issued through the Planetary Science Division of the NASA Science Mission Directorate, the National Science Foundation Grant No. AST-1238877, the University of Maryland, Eotvos Lorand University (ELTE), the Los Alamos National Laboratory, and the Gordon and Betty Moore Foundation.

\section*{DATA AVAILABILITY}
The data underlying this article are available in the article and in its online supplementary material.



\bibliographystyle{mnras}
\bibliography{example} 


\appendix

\section{Parameter correlations without  $v_\parallel$}\label{sec:expe}
\begin{figure*}
	\includegraphics[scale=0.45]{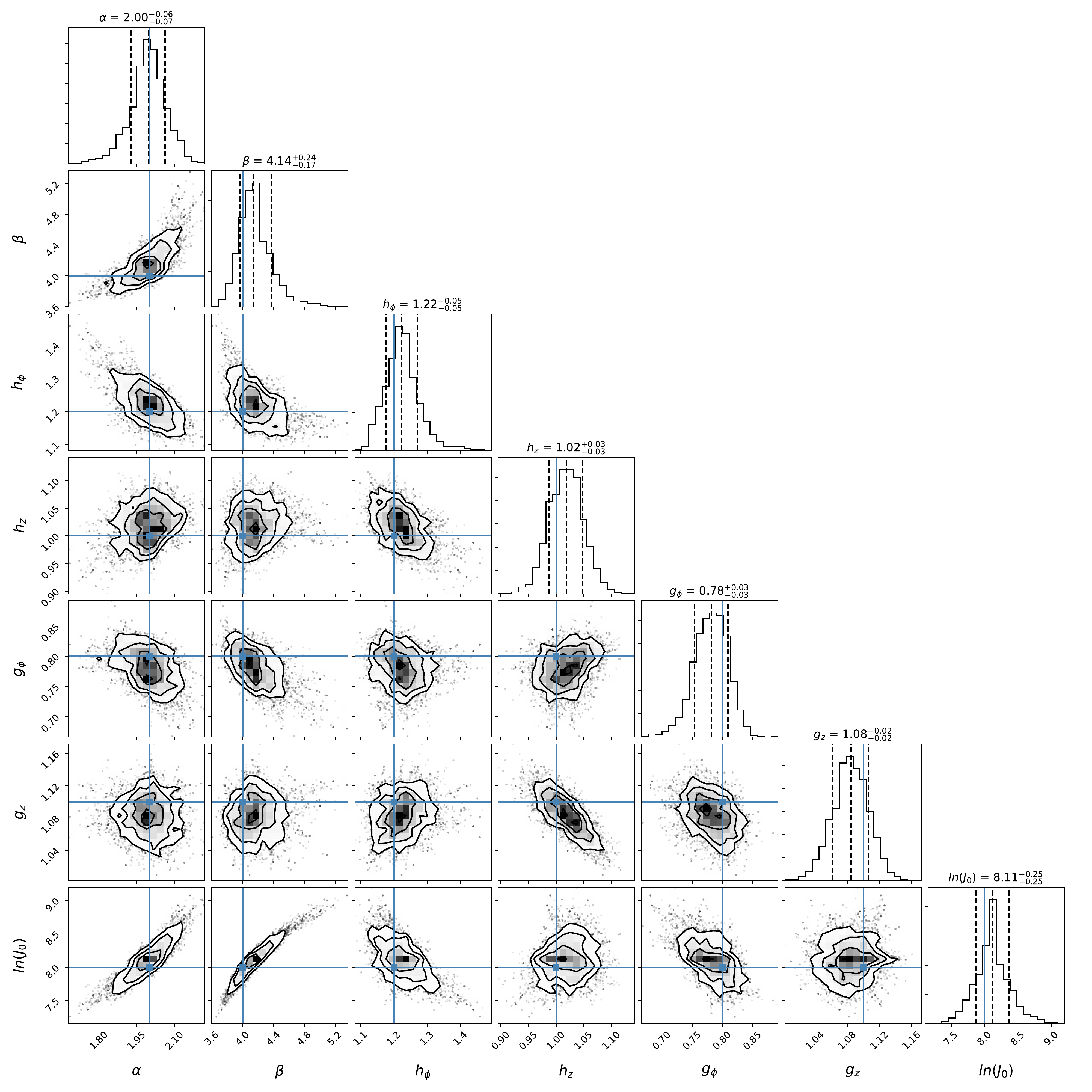}
    \caption{The posterior probability distribution of the each parameter of the mock data without $v_{los}$. The dashed lines mark represent the 1-$\sigma$ uncertainties and the cross points are the input values for the model.}
    \label{fig:mock_5d}
\end{figure*}

\end{document}